\documentclass[10pt,aps,prd,twocolumn,nofootinbib,superscriptaddress,
preprintnumbers,floatfix]{revtex4-2}

\usepackage{amsmath,amssymb,bm}
\usepackage{ifpdf}
\ifpdf
  \usepackage{graphicx}
\else
  \usepackage[dvipdfmx]{graphicx}
\fi
\graphicspath{{./}}
\DeclareGraphicsExtensions{.pdf,.png,.jpg}
\ifpdf
  \usepackage{color}
\else
  \usepackage[dvipdfmx]{color}
\fi
\ifpdf
  \usepackage{hyperref}
\else
  \usepackage[dvipdfmx]{hyperref}
\fi
\usepackage{booktabs}
\usepackage{multirow}
\numberwithin{equation}{section}

\hypersetup{colorlinks=true,linkcolor=blue,citecolor=blue,urlcolor=blue}

\newcommand{\mpl}{M_{\rm Pl}}
\newcommand{\dd}{\mathrm{d}}
\newcommand{\Omm}{\Omega_{m0}}

\newcommand{\Omr}{\Omega_{r0}}
\newcommand{\OmL}{\Omega_{\Lambda}^{(-)}}

\newcommand{\mphi}{m_{\phi}}

\begin{document}

\preprint{WUCG-26-12}

\title{Axion quintessence with a negative cosmological constant: \\Bayesian constraints from current cosmological data}

\author{Masroor C.~Pookkillath}
\email{masroorcp@gmail.com}
\affiliation{Department of Physics, Waseda University, 3-4-1 Okubo, Shinjuku, Tokyo 169-8555, Japan}

\author{Gary Shiu}
\email{shiu@physics.wisc.edu}
\affiliation{Department of Physics, University of Wisconsin-Madison, Madison, Wisconsin 53706, USA}

\author{Shinji Tsujikawa}
\email{tsujikawa@waseda.jp}
\affiliation{Department of Physics, Waseda University, 3-4-1 Okubo, Shinjuku, Tokyo 169-8555, Japan}

\date{\today}

\begin{abstract}
We constrain canonical axion quintessence with a negative cosmological constant using DES Year 5 supernovae, DESI DR2 BAO, compressed CMB information, and geometry-corrected Gold-2017 growth data. We solve the background and linear-growth equations and sample with PolyChord. Our joint prior is logarithmic in the decay constant over $0.005\leq f/M_{\rm Pl}\leq10$ and uniform in the initial angle $\theta_i=\phi_i/f\in[0,\pi]$, conditioned on $\pi-\theta_i\geq H_{\rm inf}/(2\pi f)$ with $H_{\rm inf}=10^{13}$ GeV and reduced Planck mass $M_{\rm Pl}$. The broad-prior model favors recent thawing, with median $w_{{\rm DE},0}=-0.865^{+0.068}_{-0.053}$, $m_\phi\in[0.91,4.39]\times10^{-33}$ eV, and $f/M_{\rm Pl}\in[0.589,4.899]$ (central 68\% credible intervals). Bayes factors relative to flat $\Lambda$CDM are 0.70 and 30.1 for the normalized fluctuation-conditioned prior (Convention I) and its further conditioning on numerical acceptance (Convention II). Restricting $f\leq M_{\rm Pl}$ gives $f/M_{\rm Pl}\in[0.262,0.822]$ at 68\% credibility and Bayes factors 0.233 and 32.4. Relative to the zero-vacuum axion model, the broad-prior Bayes factors are 0.620 and 1.65, providing no compelling evidence for a negative cosmological constant. Local likelihood optimization gives an Akaike information criterion difference $\Delta{\rm AIC}\simeq-10.87$ for the zero-vacuum model relative to flat $\Lambda$CDM. For this model, assuming that the instanton generating $m_\phi^2f^2=\Lambda_{\rm UV}^4e^{-S_{\rm inst}}$ satisfies the axionic Weak Gravity Conjecture, the one-sided 95\% posterior lower limit on the minimum compatible coefficient $c_{\min}$ rises from 37 to 55 as the ultraviolet scale $\Lambda_{\rm UV}$ increases from $10^{-10}M_{\rm Pl}$ to $M_{\rm Pl}$, in tension with an order-one coefficient under these assumptions.

\end{abstract}

\maketitle

\section{Introduction}
\label{sec:intro}

Type Ia supernova (SN) observations established late-time cosmic
acceleration \cite{Riess:1998cb,Perlmutter:1998np}. In the spatially flat
$\Lambda$ cold dark matter ($\Lambda$CDM) model, a positive cosmological
constant drives this acceleration. The model successfully describes
cosmic microwave background (CMB) observations
\cite{WMAP:2012nax,Planck:2018vyg}, but the physical origin of the
cosmological constant and its tiny inferred energy density remain
open theoretical questions.
Recent Dark Energy Spectroscopic Instrument (DESI) baryon acoustic
oscillation (BAO), Planck CMB, and SN results have renewed interest
in dynamical dark energy (DE)
\cite{DESI:2025bao,DESI:2025extde,DES:2025expansion}.

We define the DE equation of state as
$w_{\rm DE}\equiv p_{\rm DE}/\rho_{\rm DE}$, where $p_{\rm DE}$ and
$\rho_{\rm DE}$ are the DE pressure and energy density.
In the Chevallier--Polarski--Linder (CPL) parametrization,
$w_{\rm DE}(a)=w_0+w_a(1-a)$, with the scale factor $a$ normalized
to unity today \cite{Chevallier:2000qy,Linder:2002et}.
Combined DESI DR2 BAO, CMB, and SN data can favor evolution from
$w_{\rm DE}<-1$ at earlier times to $w_{\rm DE}>-1$ at later times.
The significance depends on the data combination
\cite{DESI:2025bao,DESI:2025extde}. 
Minimally coupled canonical quintessence
\cite{Fujii:1982,Ford:1987,Ratra:1988,Wetterich:1988,
Chiba:1997,Ferreira:1997,Copeland:1998,Caldwell:1998,Zlatev:1999,
Frieman:1995pm,Caldwell:2005tm,Copeland:2006wr}
satisfies $w_{\rm DE}\geq-1$ when the DE energy density is positive
and therefore cannot cross the phantom divide.
The possibility of phantom-divide crossing has also been explored
in scalar--tensor DE models
\cite{Ye:2024nonminimal,Wolf:2024nonminimal,Tsujikawa:2025phantom,
Wolf:2025galileon,Pookkillath:2026momentum}
and scalar--vector--tensor DE models
\cite{Tsujikawa:2026svt,Roy:2026svt}.

Thawing quintessence is initially nearly frozen at $w_{\rm DE}\simeq-1$
and evolves only at late times \cite{Caldwell:2005tm}. Models with exponential,
axion-like, and multifield potentials admit parameter ranges consistent
with cosmological observations, depending on the potential,
cosmological assumptions, and data combination
\cite{Chiba:2012cb,Tsujikawa:2013fta,Tada:2024znt,Wolf:2024xiz,
DESI:2025extde,Akrami:2025exponential,Alestas:2024curved,
Bhattacharya:2024curved,Gialamas:2025desi,Lin:2025axion,
Alestas:2025twofield}. This motivates testing explicit field
dynamics alongside phenomenological CPL fits.

The axion was introduced as a light pseudoscalar in the
Peccei--Quinn solution to the strong CP problem
\cite{Peccei:1977hh,Weinberg:1977ma,Wilczek:1977pj}. Invisible-axion
models allow the symmetry-breaking scale to lie far above the electroweak
scale \cite{Kim:1979if,Shifman:1979if}. We denote the field by $\phi$
and its decay constant by $f$. Its mass squared, $m_\phi^2$, is the
curvature of the potential at a minimum. Quantum chromodynamics (QCD)
relates the QCD axion mass to its decay constant. The axion-like
particles (ALPs) considered for DE can have much weaker explicit
symmetry breaking and masses independent of that relation
\cite{Kim:2014bcm,Marsh:2015xka}. Their approximate shift symmetry
can protect ultralight masses against large radiative corrections.

In string compactifications, axions descending from higher-dimensional
form fields can acquire nonperturbative potentials whose exponential dependence
on compactification parameters allows a broad mass spectrum \cite{Svrcek:2006yi,Arvanitaki:2009fg}.
ALPs with $m_\phi\sim H_0\sim10^{-33}\,{\rm eV}$, where $H_0$
is today's Hubble rate, are motivated DE candidates, although this
late-time target is not a universal string-theory prediction.
Axiverse DE and multi-axion dynamics are explored in
Refs.~\cite{Kamionkowski:2014zda,Katewongveerachart:2026multi,Shiu:2026bae}.

A simple realization of axion DE is provided by the periodic
pseudo-Nambu--Goldstone boson (PNGB) potential proposed in
Ref.~\cite{Frieman:1995pm}.
Related theoretical constructions invoke electroweak instantons
\cite{Nomura:2000yk,Choi:2019axion,Girmohanta:2023axion},
hidden-sector symmetry breaking \cite{Hall:2005mf},
extra dimensions \cite{Hill:2002pn}, or string-theoretic mechanisms
\cite{Choi:1999xn,KimNilles:2002,Panda:2010}.
Poly-instanton effects in type IIB compactifications have also been
proposed as a mechanism for generating the small potential scale
required for axion hilltop quintessence \cite{Caraffi:2026poly}.

Hubble friction freezes an initially misaligned, nearly stationary
axion when $H\gg m_\phi$; as $H$ decreases it can thaw toward
$w_{\rm DE}>-1$ \cite{Caldwell:2005tm,Scherrer:2007pu}.
Hilltop alignment delays rolling, so its onset depends on the
mass, decay constant, and initial displacement rather than a
universal mass threshold \cite{Dutta:2008qn,Chiba:2009sj}.

PNGB and axion quintessence have been constrained with SN, BAO,
CMB, and expansion data
\cite{Dutta:2006pn,Gupta:2011axion,SmerBarreto:2015pngb,Durrive:2018quo},
and analytic thawing descriptions tested alongside tracker and
scaling models \cite{Chiba:2012cb}. Recent work addresses
quantum-gravity-motivated hilltop models \cite{Bhattacharya:2024qg},
non-negligible initial velocities \cite{Berbig:2024twh}, and
DESI-era constraints \cite{UrenaLopez:2025ade,Lin:2025axion}.

The Chern--Simons interaction
$\mathcal L_{\phi\gamma}=-(g_{\phi\gamma}/4)\phi
F_{\mu\nu}\widetilde F^{\mu\nu}$, with electromagnetic
field strength $F_{\mu\nu}$, its dual $\widetilde F^{\mu\nu}$,
and coupling $g_{\phi\gamma}$, can produce cosmic birefringence
\cite{Harari:1992ea,Carroll:1998zi,Lue:1998mq}.
For a homogeneous field, the rotation is proportional to
$g_{\phi\gamma}\Delta\phi$ between emission and observation
\cite{Fujita:2020bwq,Fujita:2020ecn}. Minami and Komatsu reported a hint of isotropic polarization rotation
in Planck data \cite{Minami:2020odp}. Later Planck, WMAP, and
Atacama Cosmology Telescope analyses gave broadly consistent indications
\cite{DiegoPalazuelos:2022dsq,Eskilt:2022cff,
DiegoPalazuelos:2025act,Eskilt:2026actplanck}. Their interpretation
depends on foreground modeling and polarization-angle calibration.
DE explanations include electroweak axions \cite{Choi:2021biref},
monodromic quintessence \cite{Gasparotto:2022monodromy}, and
an additional vacuum component \cite{Nakagawa:2025biref}.
This motivates axion DE, although we include neither a photon
coupling nor a birefringence likelihood.

The axion shift symmetry does not fix the field-independent vacuum
energy. A negative contribution is motivated in part by anti-de Sitter
vacua in string compactifications and the challenges of constructing
controlled de Sitter vacua
\cite{Obied:2018sgi,Ooguri:2018wrx,Danielsson:2018ztv}.
Its cosmological consequences have also been investigated in
phenomenological models
\cite{Dutta:2018vmq,Visinelli:2019qqu,Calderon:2020hoc,Sen:2021wld}.
These considerations motivate testing such a contribution
without presupposing its presence.

In this work, we constrain an axion quintessence model with a negative cosmological constant, characterized by the scalar-field potential
\begin{equation}
 V(\phi)=m_\phi^2 f^2
 \left[1-\cos\left(\frac{\phi}{f}\right)\right]
 -\Lambda^{(-)},\qquad
 \Lambda^{(-)}\geq0.
 \label{eq:potential_intro}
\end{equation}
Here $\Lambda^{(-)}$ denotes the magnitude of the negative vacuum-energy
density. For $0<\Lambda^{(-)}<2m_\phi^2f^2$, the potential is positive
near its maximum but has a negative minimum.
We consider expanding cosmological solutions in which the total DE
energy density remains positive from the initial epoch to the present.

Analytic bounds on axion DE, allowing arbitrary initial displacement and velocity and either sign of the constant term, have recently been
derived beyond the slow-roll approximation \cite{Shiu:2026bae}.
Luu {\it et al.} \cite{Luu:2025ade} explored ultralight axion DE with either sign of the vacuum contribution at fixed $f=\mpl$, where $\mpl$ is the reduced Planck mass, and reported a higher peak posterior
for the negative-vacuum branch.
Such a comparison does not by itself establish a Bayesian evidence
preference, particularly once the decay constant and initial displacement are allowed to vary.
We therefore treat $f$ as a free parameter. The negative-vacuum
model $A_-$ allows $\Lambda^{(-)}$ to vary and adopts the broad
range $0.005\leq f/\mpl\leq10$. The sub-Planckian case
$A_-^{\rm sub}$ uses the same model with the restricted range
$0.005\leq f/\mpl\leq1$. The zero-vacuum model $A_0$ sets
$\Lambda^{(-)}=0$ and retains the broad prior on $f$.

Super-Planckian axion decay constants are difficult to realize
in controlled string compactifications \cite{Banks:2003sx}.
The Weak Gravity Conjecture \cite{Arkani-Hamed:2006emk},
when generalized to axions
\cite{Rudelius:2015xta,Brown:2015iha,Brown:2015lia,Heidenreich:2015wga},
constrains an instanton action-to-charge ratio.
If the constrained instanton generates the cosine potential,
the relevant bound can be written as
$f S_{\rm inst}\leq c_0\mpl$, where $S_{\rm inst}$ is its
action and $c_0$ is an order-one coefficient.

Recent work has related this bound and its coefficient to
the imaginary distance bound on Euclidean wormholes
\cite{Maldacena:2026jqd}, positivity of the gravitational path
integral \cite{DiUbaldo:2026rly}, and wormhole fragmentation
into charged defects \cite{Lin:2026ztf}.
A controlled nonperturbative expansion requires
$S_{\rm inst}\gtrsim1$; generating the DE scale through
exponential suppression of a much larger ultraviolet scale
requires $S_{\rm inst}\gg1$.
Under these assumptions, the implied restriction on $f$ is
considerably stronger than $f\leq\mpl$
\cite{Shiu:2026bae,Rudelius:2022ede}.
We therefore adopt $f\leq\mpl$ as a conservative,
theory-motivated benchmark rather than as a universal bound
on every ultraviolet completion.

Our likelihood combines DES Year 5 SN \cite{DES:2024sn},
DESI DR2 BAO \cite{DESI:2025bao}, a three-parameter compressed
CMB likelihood \cite{DESI:2025bao,Lemos:2023xhs}, and the
18-point Gold-2017 redshift-space-distortion (RSD) compilation
\cite{Nesseris:2017vor}, including its geometry correction
and WiggleZ covariance.
We solve the background and linear-growth equations for the
sampled cosmologies, determining $H_0$ through a shooting
solution of the present-day flatness condition in the axion
models, and use PolyChord \cite{Handley:2015vkr} for nested
sampling.

We compare $A_-$, $A_-^{\rm sub}$, $A_0$, and flat $\Lambda$CDM using
Bayesian evidence and the Akaike information criterion (AIC).
Evidence is normalized over the quantum-conditioned prior
(Convention I) or further conditioned on numerical acceptance
(Convention II); AIC estimates and admissible upper bounds
are distinguished in Sec.~\ref{subsec:aiccomparison}.
The refined $A_0$ solution lies near the minimum initial displacement
allowed by the quantum prior. When we extend the search to
displacements as small as $10^{-4}$ times that bound and refit the
other parameters, we find no lower $\chi^2$ within the numerical
resolution. The $A_0$ AIC estimate and the upper bounds for the
negative-vacuum models lie below the local $\Lambda$CDM reference.
The relative ranking of the axion models remains open.
The matched broad-prior $A_-/A_0$ evidence ratio tests the
negative constant; $A_-^{\rm sub}$ tests restricting $f$.
We also quantify the $A_0$ posterior's tension with the axionic
Weak Gravity Conjecture under specified instanton assumptions.

The paper is organized as follows.
Section~\ref{sec:model} defines the model, background equations, and
cosmological branch considered.
Section~\ref{sec:phenomenology} discusses thawing dynamics and the role
of the negative constant.
Section~\ref{sec:method} describes the observables, likelihoods,
numerical methods, priors, and evidence normalization.
Section~\ref{sec:results} presents parameter constraints, the
reconstructed equation of state, model comparisons, and the
sub-Planckian prior test.
Section~\ref{sec:conclusion} concludes.
We use natural units $c=\hbar=1$, except where $c$ is explicitly restored
in cosmological distance definitions.

\section{Axion quintessence with a negative vacuum energy}
\label{sec:model}

We describe the axion field $\phi$ by the action
\begin{align}
 S={}&\int \dd^4x\sqrt{-g}\left[
 \frac{\mpl^2}{2}R-\frac12 g^{\mu\nu}
 \nabla_\mu\phi\nabla_\nu\phi-V(\phi)\right]\nonumber\\
 &+S_m[g_{\mu\nu},\Psi_m],
 \label{eq:action}
\end{align}
where $g_{\mu\nu}$ is the metric, $g$ its determinant, $R$
the Ricci scalar, $\nabla_\mu$ the metric-compatible covariant
derivative, and $\mpl$ the reduced Planck mass. The matter action
$S_m$ describes radiation, baryons, and CDM, whose
degrees of freedom are collectively denoted by $\Psi_m$.
We rewrite the potential in Eq.~\eqref{eq:potential_intro} as
\begin{equation}
 V(\phi)=U(\phi)-\Lambda^{(-)},\qquad
 U(\phi)=\mphi^2f^2\left[1-\cos\left(\frac{\phi}{f}\right)\right].
 \label{eq:potential}
\end{equation}
Since the constant term is independent of $\phi$, the scalar force
is determined by $V_{,\phi}=U_{,\phi}=\mphi^2f\sin(\phi/f)$.

We consider a spatially flat Friedmann--Lema\^itre--Robertson--Walker
(FLRW) spacetime described by the line element
\begin{equation}
 \dd s^2=-\dd t^2+a^2(t)\delta_{ij}\dd x^i\dd x^j,
 \label{eq:flrwmetric}
\end{equation}
where $t$ is cosmic time, $a(t)$ is the scale factor normalized
to unity today, $x^i$ are comoving Cartesian coordinates,
and $\delta_{ij}$ is the Kronecker delta.
We take the scalar field to be spatially homogeneous,
$\phi=\phi(t)$. The Hubble expansion rate is defined by
$H\equiv\dot a/a$, where an overdot denotes differentiation
with respect to $t$. 

The axion energy density and pressure, excluding the constant term, are
\begin{equation}
 \rho_a=\frac{\dot\phi^2}{2}+U(\phi),\qquad
 p_a=\frac{\dot\phi^2}{2}-U(\phi),
 \label{eq:axionrho}
\end{equation}
while inclusion of the constant term gives the total DE quantities
\begin{equation}
 \rho_{\rm DE}=\rho_a-\Lambda^{(-)},\qquad
 p_{\rm DE}=p_a+\Lambda^{(-)}.
 \label{eq:derhop}
\end{equation}
We denote the energy densities of radiation, baryons, and
CDM by $\rho_r$, $\rho_b$, and $\rho_c$,
respectively, and define the total nonrelativistic matter
density as $\rho_m\equiv\rho_b+\rho_c$.
Radiation and nonrelativistic matter are separately conserved, with their continuity equations given by
\begin{equation}
 \dot\rho_r+4H\rho_r=0,\qquad
 \dot\rho_m+3H\rho_m=0.
 \label{eq:fluid_cons}
\end{equation}
Variation of the action yields the background equations
\begin{align}
 3\mpl^2H^2&=\rho_r+\rho_m+\rho_{\rm DE},
 \label{eq:friedmann1}\\
 -2\mpl^2\dot H&=\rho_m+\frac{4}{3}\rho_r+\dot\phi^2,
 \label{eq:friedmann2}\\
 \ddot\phi+3H\dot\phi
 &+\mphi^2f\sin\left(\frac{\phi}{f}\right)=0.
 \label{eq:kgeq}
\end{align}
The constant contributes to Eq.~\eqref{eq:friedmann1} but
cancels from $\rho_{\rm DE}+p_{\rm DE}=\dot\phi^2$ and
leaves $V_{,\phi}$ unchanged. The scalar equation gives
\begin{equation}
 \dot\rho_{\rm DE}+3H(\rho_{\rm DE}+p_{\rm DE})
 =\dot\rho_{\rm DE}+3H\dot\phi^2=0.
 \label{eq:deconservation}
\end{equation}
Differentiating the first Friedmann equation and using
Eqs.~\eqref{eq:fluid_cons} and \eqref{eq:deconservation} gives the
second Friedmann equation on the expanding branch. On this branch
$\dot\rho_{\rm DE}=-3H\dot\phi^2\leq0$, so the DE density is
nonincreasing with cosmic time.

To distinguish the axion contribution from the complete DE sector,
we define
\begin{equation}
 w_\phi\equiv\frac{p_a}{\rho_a},\qquad
 \Omega_\phi\equiv\frac{\rho_a}{3\mpl^2H^2},
 \label{eq:axion_w_omega}
\end{equation}
for the axion alone, and
\begin{equation}
 w_{\rm DE}\equiv\frac{p_{\rm DE}}{\rho_{\rm DE}},\qquad
 \Omega_{\rm DE}\equiv\frac{\rho_{\rm DE}}{3\mpl^2H^2},
 \label{eq:wde_def}
\end{equation}
for the total DE component. We also use
$\Omega_m\equiv\rho_m/(3\mpl^2H^2)$ and
$\Omega_r\equiv\rho_r/(3\mpl^2H^2)$, so spatial flatness gives
$\Omega_m+\Omega_r+\Omega_{\rm DE}=1$.
Writing $H_0$ for today's Hubble rate and using a subscript $0$ for
present-day values, the dimensionless magnitude of the negative
vacuum energy is
\begin{equation}
 \OmL\equiv\frac{\Lambda^{(-)}}{3\mpl^2H_0^2}\geq0.
 \label{eq:OmegaLambdaMinus}
\end{equation}
These definitions imply
$\Omega_{{\rm DE},0}=\Omega_{\phi,0}-\OmL$.
Using $\rho_a-p_a=2U(\phi)$ and
$\rho_{\rm DE}-p_{\rm DE}=2[U(\phi)-\Lambda^{(-)}]$,
we also obtain the exact identity
\begin{equation}
 (1-w_{\phi,0})\Omega_{\phi,0}
 =(1-w_{{\rm DE},0})\Omega_{{\rm DE},0}+2\OmL.
 \label{eq:wphi_wde_relation}
\end{equation}
Moreover, $\rho_{\rm DE}+p_{\rm DE}=\dot\phi^2$ implies
\begin{equation}
 1+w_{\rm DE}=\frac{\dot\phi^2}{\rho_{\rm DE}}\geq0
 \qquad (\rho_{\rm DE}>0).
 \label{eq:nonnull}
\end{equation}
Hence $w_{\rm DE}$ can reach $-1$ when $\dot\phi=0$ but cannot
cross to $w_{\rm DE}<-1$ while $\rho_{\rm DE}$ remains positive.
At $\rho_{\rm DE}=0$, this equation-of-state ratio is undefined and
may diverge without implying a spacetime singularity or a ghost.
Our inference is restricted to expanding backgrounds with
$\rho_{\rm DE}>0$ from the initial epoch to the present; this is a
choice of cosmological branch rather than a general prohibition of
negative DE density.

For numerical evolution we introduce the e-folding number
$N\equiv\ln a=-\ln(1+z)$, where $z$ is redshift, and
$E\equiv H/H_0$. A subscript $N$ denotes differentiation with respect to $N$, so $\dot\phi=H\phi_N$. 
The Klein--Gordon and
second Friedmann equations take the form
\begin{eqnarray}
\hspace{-0.7cm}
& &
 \phi_{NN}+\left(3+\frac{H_N}{H}\right)\phi_N
 +\frac{\mphi^2f}{H^2}\sin\left(\frac{\phi}{f}\right)=0,
 \label{eq:kg_N}\\
\hspace{-0.7cm}
& &
 \frac{H_N}{H}=-\frac{3}{2}\Omega_m-2\Omega_r
 -\frac{\phi_N^2}{2\mpl^2}.
 \label{eq:HprimeH_N}
\end{eqnarray}
The first Friedmann equation (\ref{eq:friedmann1}) 
gives the algebraic relation
\begin{align}
 E^2\left(1-\frac{\phi_N^2}{6\mpl^2}\right)
 ={}&\Omr e^{-4N}+\Omm e^{-3N}\nonumber\\
 &+\frac{U(\phi)}{3\mpl^2H_0^2}-\OmL,
 \label{eq:EN_alg}
\end{align}
where $\Omm$ and $\Omr$ are the present matter and radiation
density parameters. The factor $1-\phi_N^2/(6\mpl^2)$ is a
background denominator arising from the Friedmann constraint, not a
kinetic coefficient in the action. As detailed in
Sec.~\ref{subsec:priors}, we select the branch on which this factor
is positive.

The change of variables $\theta\equiv\phi/f$ provides an
alternative description of the scalar dynamics. In terms of $\theta$,
Eq.~\eqref{eq:kg_N} becomes
\begin{equation}
 \theta_{NN}+\left(3+\frac{H_N}{H}\right)\theta_N
 +\frac{\mphi^2}{H^2}\sin\theta=0.
 \label{eq:theta_N}
\end{equation}
Although $f$ cancels from the explicit force term, it enters
$H$ through $\phi_N^2=f^2\theta_N^2$ and
$U(f\theta)=m_\phi^2f^2(1-\cos\theta)$ in
Eq.~\eqref{eq:EN_alg}; the initial angle alone does not specify
the coupled evolution.

To describe thawing solutions, we initialize the scalar field at
redshift $z_i=10^7$ according to
\begin{equation}
 \phi(N_i)=f\theta_i,\qquad \phi_N(N_i)=0,
 \qquad N_i=-\ln(1+z_i),
 \label{eq:initial_conditions}
\end{equation}
where $\theta_i=\phi_i/f\in[0,\pi]$ and $\phi_i=\phi(N_i)$.
Equivalently, $\theta(N_i)=\theta_i$ and $\theta_N(N_i)=0$.
Zero velocity approximates a frozen field for $H(z_i)\gg m_\phi$
but need not be the exact regular solution at finite $z_i$.
We reconstruct $w_{\rm DE}$ and $\Omega_{\rm DE}$ from
Eqs.~\eqref{eq:derhop} and \eqref{eq:wde_def} using
$\dot\phi=H\phi_N$.

The periodic axion potential is bounded by
$0\leq U(\phi)\leq 2\mphi^2f^2$. Hence, the total
potential $V(\phi)=U(\phi)-\Lambda^{(-)}$ can be positive
only if
\begin{equation}
 \Lambda^{(-)}<2\mphi^2f^2.
 \label{eq:hilltopsupport}
\end{equation}
Ref.~\cite{Shiu:2026bae} discusses the corresponding
condition for generalized periodic potentials.
This inequality ensures that a region with $V(\phi)>0$
exists, but does not guarantee a cosmologically viable solution. With the zero-velocity initial condition
\eqref{eq:initial_conditions}, the chosen initial angle must
also satisfy $U(\phi_i)>\Lambda^{(-)}$ to start on the
positive-DE-density branch.
Additional conditions imposed in our statistical
analysis are described in Sec.~\ref{subsec:priors}.

\section{Thawing dynamics and the role of negative vacuum energy}
\label{sec:phenomenology}

Equation~\eqref{eq:kgeq} can be written as
$(a^3\dot\phi)^{\cdot}=-a^3U_{,\phi}$.
Integrating this relation with the initial condition
$\dot\phi(t_i)=0$, we obtain the exact scalar-field velocity,
\begin{equation}
 \dot\phi(t)=-\frac{1}{a^3(t)}\int_{t_i}^{t}
 a^3(\tilde t)U_{,\phi}[\phi(\tilde t)]\,\dd\tilde t.
 \label{eq:forceintegral}
\end{equation}
The factor $[a(\tilde t)/a(t)]^3$ describes how cosmic expansion
damps the velocity generated by the potential force $-U_{,\phi}$.
For $0<\theta_i<\pi$, we have $U_{,\phi}(\phi_i)>0$, so the
field initially moves toward smaller $\phi$.
Appreciable rolling can develop as $H$ decreases relative to $m_\phi$.

If the scalar-field acceleration is negligible compared with
the Hubble friction term, $|\ddot\phi|\ll 3H|\dot\phi|$,
the scalar-field equation yields
\begin{equation}
 \dot\phi\simeq-\frac{U_{,\phi}}{3H},\qquad
 1+w_{\rm DE}\simeq
 \frac{U_{,\phi}^2}{9H^2\rho_{\rm DE}}.
 \label{eq:thawestimate}
\end{equation}
However, $w_{\rm DE}\simeq-1$ does not imply that the
scalar-field acceleration is negligible during thawing
\cite{Shiu:2026bae}. For a dominant fluid with approximately
constant $w_{\rm bg}>-1$, $H\simeq2/[3(1+w_{\rm bg})t]$,
and slowly varying $U_{,\phi}$, Eq.~\eqref{eq:forceintegral}
gives, after the decaying initial-condition term becomes negligible,
\begin{equation}
 \dot\phi\simeq-\frac{2}{3(3+w_{\rm bg})}
 \frac{U_{,\phi}}{H}.
 \label{eq:fluid_thawing}
\end{equation}
The coefficient of $U_{,\phi}/H$ in Eq.~\eqref{eq:fluid_thawing}
is $-2/9$ during matter domination ($w_{\rm bg}=0$) and $-1/5$
during radiation domination ($w_{\rm bg}=1/3$). Neglecting $\ddot\phi$
in the field equation would give $-1/3$ in both cases, as in
Eq.~\eqref{eq:thawestimate}. In fact,
$|\ddot\phi|/(3H|\dot\phi|)\simeq(1+w_{\rm bg})/2$, which is
$1/2$ and $2/3$ in these two regimes, respectively. Our numerical analysis solves the full
Eqs.~\eqref{eq:kg_N}--\eqref{eq:EN_alg} without these approximations.

At fixed $m_\phi$ and $f$, the initial angle fixes the potential slope
$U_{,\phi}=m_\phi^2f\sin(\phi/f)$ and curvature
$U_{,\phi\phi}=m_\phi^2\cos(\phi/f)$ at the starting point.
Near the hilltop the driving force $-U_{,\phi}$ is small,
but $U_{,\phi\phi}\simeq-m_\phi^2<0$. Small homogeneous
displacements from the maximum are unstable, although
Hubble friction can delay their growth until late times.
Thawing thus depends on both the potential and the expansion history.

The negative vacuum contribution changes the expansion history
without changing the potential slope or curvature. At fixed
$m_\phi$, $f$, $\phi$, $\dot\phi$, and matter and radiation
densities, increasing $\Lambda^{(-)}$ lowers
$\rho_{\rm DE}=\rho_a-\Lambda^{(-)}$ and hence $H^2$,
reducing Hubble friction. On the positive-density branch,
it also increases $1+w_{\rm DE}=\dot\phi^2/\rho_{\rm DE}$
at fixed $\dot\phi\ne0$ through the smaller denominator.
These fixed-state effects differ from comparisons of solutions
with the same present DE density: the latter generally require
readjusting axion parameters and can change $\dot\phi$ itself.

For the illustrative histories, we fix $H_0=67\,{\rm km\,s^{-1}\,Mpc^{-1}}$,
$\Omega_{m0}=0.32$, and $\Omega_{r0}=9.31924\times10^{-5}$,
giving $\Omega_{{\rm DE},0}=0.679907$.
For each $(m_\phi/H_0,\theta_i,\OmL)$, we integrate
Eqs.~\eqref{eq:kg_N}--\eqref{eq:EN_alg} with
Eq.~\eqref{eq:initial_conditions}, adjusting $f$ until
$E^2(0)=1$. Since $\Omega_{{\rm DE},0}$ is fixed, increasing
$\OmL$ requires a larger axion density,
$\Omega_{\phi,0}=\Omega_{{\rm DE},0}+\OmL$.
The flatness condition thus links the chosen negative vacuum term
to the value of $f$ used in each illustration. In the statistical
analysis we sample $f$ and determine $H_0$ from flatness
(Sec.~\ref{subsec:numericalsolver}).

\begin{figure}[tbp]
 \centering
 \includegraphics[width=0.95\linewidth]{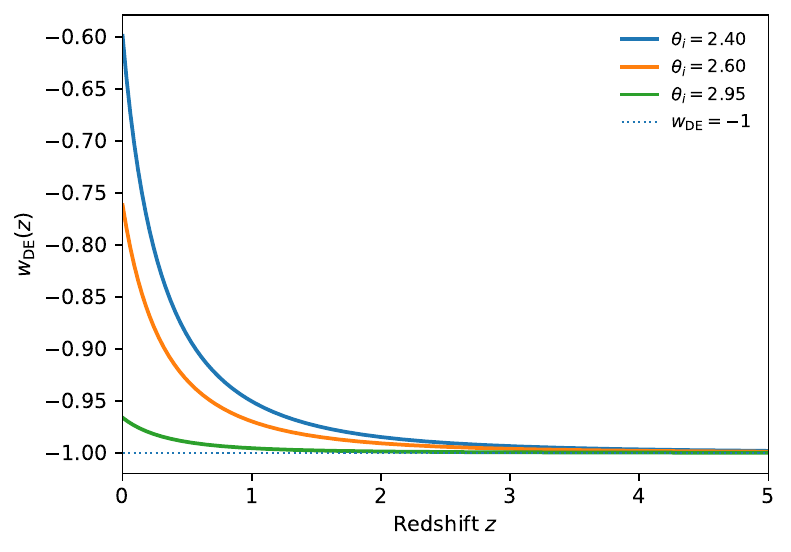}
 \caption{Normalized illustrative histories for $\theta_i=2.40,2.60,2.95$, at $m_\phi/H_0=2$ and $\OmL=1.6$. In the same order, the flatness solutions give $f/\mpl\simeq1.060,0.997,0.934$ and the resulting present equations of state are $w_{{\rm DE},0}=-0.599,-0.762,-0.966$. Moving closer to the hilltop suppresses thawing in this family.}
 \label{fig:fig1_phi_i}
\end{figure}

\begin{figure}[tbp]
 \centering
 \includegraphics[width=0.95\linewidth]{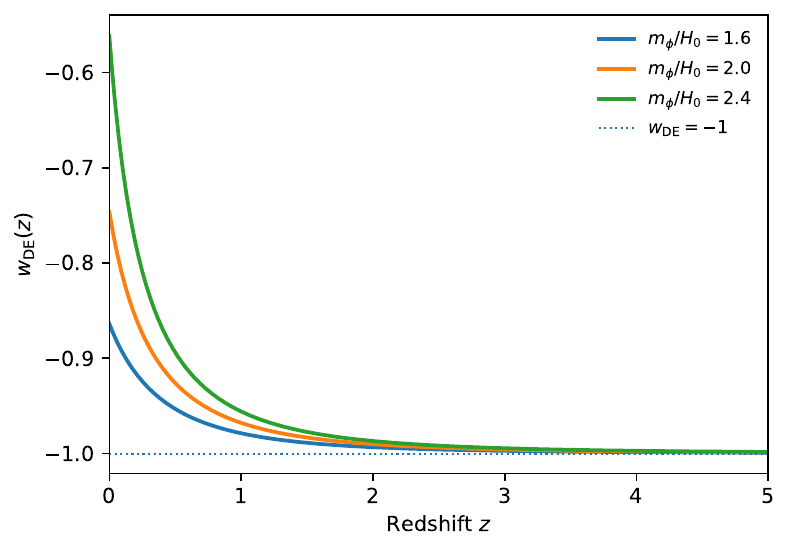}
 \caption{Normalized illustrative histories for $m_\phi/H_0=1.6,2.0,2.4$, with $\theta_i=2.58$ and $\OmL=1.6$. The corresponding flatness solutions are $f/\mpl\simeq1.232,1.002,0.855$, giving $w_{{\rm DE},0}=-0.864,-0.746,-0.562$. The mass ratio is varied while $f$ is readjusted; this is not a scan at fixed decay constant.}
 \label{fig:fig2_mphi}
\end{figure}

Figures~\ref{fig:fig1_phi_i} and \ref{fig:fig2_mphi} 
show how the
underlying axion dynamics controls the departure from
$w_{\rm DE}=-1$ at fixed $\OmL=1.6$.
In Fig.~\ref{fig:fig1_phi_i}, moving the initial field closer
to the hilltop reduces its force and delays rolling, leading to
$w_{{\rm DE},0}$ closer to $-1$ along the normalized family.
Figure~\ref{fig:fig2_mphi} shows a larger present-day departure
from $-1$ as $\mphi/H_0$ increases at fixed $\theta_i$.
Because we readjust $f$ in each case, the curves describe a family
of normalized solutions in which more than one parameter varies.

%%%%%%%%%%%%%%%%%%%%%%%
%%%%%%%%%%%%%%%%%%%%%%%

%%%%%%%%%%%%%%%%%%%%%%%
%%%%%%%%%%%%%%%%%%%%%%%%

\begin{figure}[tbp]
 \centering
 \includegraphics[width=0.95\linewidth]{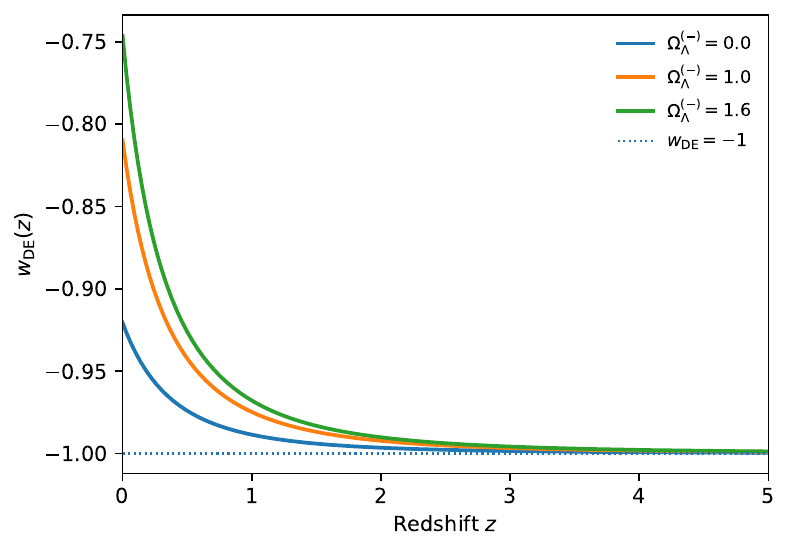}
 \caption{Normalized illustrative histories for $\OmL=0,1.0,1.6$, with $m_\phi/H_0=2$ and $\theta_i=2.58$. Flatness requires $f/\mpl\simeq0.548,0.861,1.002$, respectively. For this explicitly defined family, the present equation of state changes monotonically from $-0.9206$ to $-0.8099$ and $-0.7464$.}
 \label{fig:fig3_oml}
\end{figure}

Figure~\ref{fig:fig3_oml} varies $\OmL=0,1.0,1.6$ at
$m_\phi/H_0=2$ and $\theta_i=2.58$, readjusting $f$ at
fixed $H_0$ and $\rho_{{\rm DE},0}$.
For each member of this family, the axion supplies the density
$\rho_{a0}=3\mpl^2H_0^2(\Omega_{{\rm DE},0}+\OmL)$.
The present scalar kinetic energy $\dot\phi_0^2/2$ increases
along the sequence. Since $\rho_{{\rm DE},0}$ is fixed,
Eq.~\eqref{eq:nonnull} attributes the resulting values
$w_{{\rm DE},0}=-0.9206,-0.8099,-0.7464$ to the change in
$\dot\phi_0^2$. This trend depends on how the solutions were
normalized.
Ref.~\cite{Shiu:2026bae} derives a lower bound on the time
required for a given increase in $\epsilon=-\dot H/H^2$.
A negative constant reduces this bound when the other quantities
entering it are held fixed. Connecting this bound to the actual
histories in Fig.~\ref{fig:fig3_oml} requires the full background,
since $\epsilon$ includes matter and radiation
[Eq.~\eqref{eq:HprimeH_N}].

%%%%%%%%%%%%%%%%%%%%%%%%
%%%%%%%%%%%%%%%%%%%%%%%%%

\begin{figure}[tbp]
 \centering
 \includegraphics[width=0.95\linewidth]{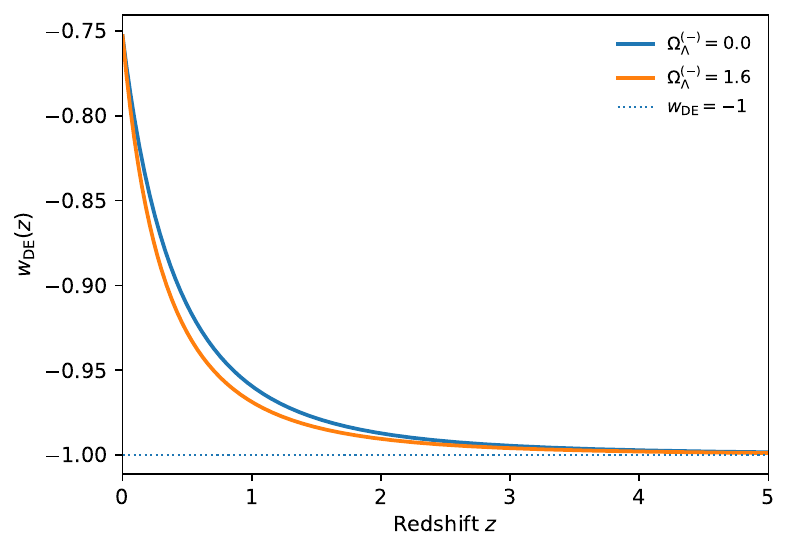}
 \caption{Two normalized illustrative histories with $w_{{\rm DE},0}=-0.753$ and $m_\phi/H_0=2$. The $\OmL=0$ solution has $(f/\mpl,\theta_i)\simeq(0.690,2.006)$, and the $\OmL=1.6$ solution has $(1.000,2.589)$. Both share the present matter, radiation, and net DE densities specified in the text. Their different shapes illustrate the additional freedom beyond a single present-day equation-of-state value.}
 \label{fig:fig4_compare}
\end{figure}

A complementary comparison is shown in Fig.~\ref{fig:fig4_compare},
where the two backgrounds have the same $\rho_{{\rm DE},0}$ and
$w_{{\rm DE},0}$, but different values of $\OmL$.
Defining $K_0\equiv\dot\phi_0^2/2$, the relations in
Sec.~\ref{sec:model} imply
\begin{align}
 K_0&=\frac{1+w_{{\rm DE},0}}{2}\rho_{{\rm DE},0},\\
 U(\phi_0)&=\frac{1-w_{{\rm DE},0}}{2}\rho_{{\rm DE},0}
 +\Lambda^{(-)}.
 \label{eq:matchedkinetic}
\end{align}
Fixing $\rho_{{\rm DE},0}$ and $w_{{\rm DE},0}$ therefore fixes
$K_0$, while a larger negative constant requires larger $U(\phi_0)$.
The present density and equation of state alone leave
the potential slope and curvature undetermined. The two solutions
in Fig.~\ref{fig:fig4_compare} have the same $w_{{\rm DE},0}$
but different $w_{\rm DE}(z)$ at earlier times. They also differ
in $f$ and $\theta_i$, illustrating the degeneracy among those
parameters and $\Lambda^{(-)}$.

%%%%%%%%%%%%%%%%%%%%%%%%%
%%%%%%%%%%%%%%%%%%%%%%%%%

The illustrative values of $f$ are not restricted by the
alternative $f\leq\mpl$ prior. Observational constraints on
these degeneracies follow in Secs.~\ref{sec:method} and
\ref{sec:results}.

%%%%%%%%%%%%%%%%%%%%%%%%%
\section{Numerical and statistical methodology}
\label{sec:method}
%%%%%%%%%%%%%%%%%%%%%%%%%

We solve the background and linear-growth equations for the
distance and growth observables entering our four likelihoods.
PolyChord \cite{Handley:2015vkr}, interfaced with \textsc{Cobaya}
\cite{Cobaya:2020}, supplies weighted posteriors and evidences;
GetDist \cite{Lewis:2019xzd} summarizes the posteriors.
All models use the same likelihoods and calibration bounds,
with additional acceptance criteria specified below.

\subsection{Distances, CMB quantities, and linear growth}
\label{subsec:observables}

For the spatially flat FLRW background introduced in
Sec.~\ref{sec:model}, the transverse comoving distance
$D_M(z)$, angular-diameter distance $D_A(z)$, and
luminosity distance $D_L(z)$ are given by
\begin{align}
 D_M(z)&=c\int_0^z\frac{\dd\tilde z}{H(\tilde z)},
 \nonumber\\
 D_A(z)&=\frac{D_M(z)}{1+z},\qquad
 D_L(z)=(1+z)D_M(z),
 \label{eq:distances}
\end{align}
where $c$ is the speed of light,
restored here in the distance expressions.
We also define the Hubble distance $D_H(z)$ and the
volume-averaged BAO distance $D_V(z)$ as
\begin{equation}
 D_H(z)=\frac{c}{H(z)},\qquad
 D_V(z)=\left[zD_H(z)D_M^2(z)\right]^{1/3}.
 \label{eq:DV}
\end{equation}
The DESI DR2 BAO data constrain $D_M/r_{\rm drag}$ and
$D_H/r_{\rm drag}$ through anisotropic measurements,
or $D_V/r_{\rm drag}$ through isotropic measurements.
Here $r_{\rm drag}$ denotes the comoving sound horizon
at the baryon-drag epoch.

For the compressed CMB likelihood, we define
$h\equiv H_0/(100\,{\rm km\,s^{-1}\,Mpc^{-1}})$ and
the physical baryon and CDM densities
$\omega_b\equiv\Omega_{b0}h^2$ and
$\omega_c\equiv\Omega_{c0}h^2$, respectively.
We also introduce $\omega_{bc}\equiv\omega_b+\omega_c$,
which coincides with the total nonrelativistic matter
density $\omega_m\equiv\Omega_{m0}h^2$ in the
massless-neutrino approximation adopted here.
The compressed CMB data vector is
\begin{equation}
 \boldsymbol d_{\rm CMB}=(\theta_\star,\omega_b,\omega_{bc}),
 \qquad
 \theta_\star\equiv\frac{r_\star}{D_M(z_\star)},
 \label{eq:cmbvector}
\end{equation}
where $z_\star$ is the photon-decoupling redshift
and $r_\star$ is the comoving sound horizon at that
epoch. If massive neutrinos were included,
$\omega_{bc}$ would remain the baryon-plus-CDM
density, while $\omega_m$ would also contain the
nonrelativistic neutrino contribution.

The RSD prediction uses the total matter contrast
$\delta_m=(\delta\rho_b+\delta\rho_c)/\rho_m$,
where $\delta\rho_b$ and $\delta\rho_c$ are the baryon and
CDM density perturbations.

For the canonical axion, the rest-frame sound speed squared is
$c_\phi^2=1$. We adopt the quasistatic approximation on scales
well inside its sound horizon, $k^2/a^2\gg H^2$, where $k$
is the comoving wavenumber. In this regime, spatial gradients
dominate over time derivatives of the metric and scalar
perturbations, provided freely oscillating modes are subdominant
\cite{Boisseau:2000qs,Tsujikawa:2007qs,DeFelice:2011qs,Kase:2018qs}.
The potential curvature obeys $|U_{,\phi\phi}|\leq\mphi^2$.
We additionally require $k^2/a^2\gg|U_{,\phi\phi}|$ for the
spatial-gradient term to dominate the scalar effective-mass
term. This hierarchy is not guaranteed throughout the mass
prior, whose upper limit is $10^{-30}\,{\rm eV}$. The smooth-DE
growth equation below is therefore an explicit likelihood
approximation; its applicability to a given solution depends
on both the curvature and the wavenumbers probed by the RSD data.
Under these assumptions, we neglect axion density perturbations
at leading order.
Neglecting radiation perturbations as well, the linearized
Einstein and matter equations in this quasistatic, smooth-DE
limit yield
\begin{equation}
 \delta_m''+\left(2+\frac{H'}{H}\right)\delta_m'
 -\frac{3}{2}\Omega_m\delta_m=0.
 \label{eq:growth_approx}
\end{equation}
Here a prime denotes differentiation with respect to $N=\ln a$.
Equation~\eqref{eq:growth_approx} retains the time evolution of
the matter contrast but is not a full Boltzmann treatment
of cosmological perturbations.

We define the logarithmic growth rate
$f_g\equiv\dd\ln\delta_m/\dd\ln a$, distinguished
from the axion decay constant $f$.
We also introduce $\sigma_8\equiv\sigma_{8,0}$,
the present-day rms linear matter density fluctuation
smoothed over spheres of radius $8h^{-1}\,{\rm Mpc}$.
In the scale-independent growth approximation,
the RSD observable is given by
\begin{equation}
 f_g(z)\sigma_8(z)=\sigma_8
 \frac{\delta_m'[N(z)]}{\delta_m(0)},
 \qquad N(z)=-\ln(1+z).
 \label{eq:fs8}
\end{equation}
The normalization of $\delta_m$ cancels in this
expression, and $\sigma_8$ is treated as a sampled
amplitude parameter. We also define
$S_8\equiv\sigma_8\sqrt{\Omega_{m0}/0.3}$.

\subsection{Parameters and admissible backgrounds}
\label{subsec:priors}

Table~\ref{tab:axion_models} summarizes the three axion priors.
Relative to $A_-$, $A_-^{\rm sub}$ lowers the upper bound on
$f$ to $\mpl$, while $A_0$ fixes $\OmL=0$; all other
underlying bounds are retained.

\begin{table}[t]
\caption{The three axion cases and their defining prior bounds.
The ranges shown are the underlying bounds before conditioning
on Eq.~\eqref{eq:quantum_cut}. Table~\ref{tab:axion_priors} lists
the underlying priors for the broad-prior case $A_-$.}
\label{tab:axion_models}
\begin{ruledtabular}
\begin{tabular}{lccc}
Model & $f/\mpl$ & $\OmL$ & Free parameters \\
\hline
$A_-$ & $[0.005,10]$ & $[0,3]$ & 7 \\
$A_-^{\rm sub}$ & $[0.005,1]$ & $[0,3]$ & 7 \\
$A_0$ & $[0.005,10]$ & $0$ (fixed) & 6 \\
\end{tabular}
\end{ruledtabular}
\end{table}

The physical parameter basis for $A_-$ is
\begin{equation}
 \boldsymbol\vartheta_{A_-}=\left(
 \omega_b,\omega_c,\log_{10}\frac{\mphi}{{\rm eV}},
 \log_{10}\frac{f}{\mpl},\OmL,\theta_i,\sigma_8\right).
 \label{eq:param_basis}
\end{equation}
Table~\ref{tab:axion_priors} specifies the independent top-hat
densities before conditioning on the inflationary benchmark.
Compared with the narrower parameter domain often used for axion
DE, we allow $f/\mpl$ down to $0.005$ and $m_\phi$ up to
$10^{-30}\,{\rm eV}$. The prior is logarithmic in both $f$ and
$m_\phi$ and uniform in the initial angle, rather than logarithmic
in its displacement from the hilltop.

The lower cutoff $f/\mpl=0.005$ is also chosen to sample values
below the estimate $f\lesssim0.008\mpl$ discussed for axion early
DE in Ref.~\cite{Rudelius:2022ede}.
This estimate relies on the axion weak gravity conjecture and
specific choices of the axion mass and ultraviolet scale
$\Lambda_{\rm UV}$, and admits possible loopholes.
We use it to motivate prior coverage, rather than as a universal
bound on late-time quintessence. Extending the underlying prior
to these small decay constants does not imply that the same
region contains accepted late-time solutions. Initial conditions must 
satisfy the quantum cut introduced below
and yield the required present-day DE density after evolution
from the hilltop (Sec.~\ref{subsec:subplanck}).

During inflation with Hubble rate $H_{\rm inf}$, an effectively
light spectator field acquires a characteristic fluctuation
$H_{\rm inf}/(2\pi)$ per logarithmic interval in scale
\cite{StarobinskyYokoyama:1994}.
Our background analysis follows homogeneous classical trajectories
initialized on the $\phi_i<\pi f$ side of the maximum.
If the mean displacement from the maximum is smaller than
this fluctuation scale, quantum diffusion during inflation
can carry local field values across the hilltop into $\phi>\pi f$.
We use this scale to define a benchmark for the classical
initial displacement,
\begin{align}
 \delta_i&\equiv\pi-\theta_i,\qquad
 \Delta\phi_i\equiv\pi f-\phi_i=f\delta_i,\nonumber\\
 Q:\quad\delta_i&\geq\delta_{\min}(f)
 \equiv\frac{H_{\rm inf}}{2\pi f}.
 \label{eq:quantum_cut}
\end{align}
We thus require $\Delta\phi_i\geq H_{\rm inf}/(2\pi)$
for the initial homogeneous background, including equality at
the cut. This benchmark uses the fluctuation in one logarithmic
interval of scale. Individual field realizations can still cross
the maximum stochastically. The cumulative displacement depends
on the inflationary history and the coarse-graining scale.
Our prior uses Eq.~\eqref{eq:quantum_cut} directly.

The scale $H_{\rm inf}=10^{13}\,{\rm GeV}$ is chosen as a
representative plateau-inflation benchmark. For a canonical
single inflaton in Einstein gravity, with adiabatic perturbations
generated during slow roll,
\begin{align}
 A_s&\simeq\frac{H_*^2}{8\pi^2\epsilon_*\mpl^2},\qquad
 \epsilon_*\equiv-\frac{\dot H_*}{H_*^2},\nonumber\\
 \epsilon_*&\simeq1.02\times10^{-4}
 \left(\frac{H_*}{10^{13}\,{\rm GeV}}\right)^2
 \left(\frac{2.1\times10^{-9}}{A_s}\right).
 \label{eq:inflation_scale_epsilon}
\end{align}
Here $A_s$ is the dimensionless primordial curvature-power amplitude;
inflationary quantities with subscript $*$ are evaluated at pivot-mode
Hubble exit, $k_*=a_*H_*$. With $A_s\simeq2.1\times10^{-9}$
\cite{Planck:2018vyg,Planck:2018inflation}, the adopted scale
corresponds to $\epsilon_*\simeq10^{-4}$. 
For comparison, in the Einstein frame, the Starobinsky model
\cite{Starobinsky:1980} gives
$\epsilon_*\simeq3/(4\mathcal N_*^2)$ and
$H_*\simeq(1.4$--$1.7)\times10^{13}\,{\rm GeV}$
for $\mathcal N_*=50$--$60$ $e$-folds before the end of inflation.
Our rounded scale is thus a representative plateau benchmark,
rather than an exact prediction at a specified $\mathcal N_*$.
Lower-scale inflation, including models with an additional exit
sector \cite{RossGerman:2010}, remains possible. Lowering
$H_{\rm inf}$ lowers $\delta_{\min}$ and admits closer hilltop
alignment, thereby changing the allowed domain and its normalized
prior. 

The adopted value $H_{\rm inf}=10^{13}\,{\rm GeV}$ is a benchmark
assumption, not an observational lower bound.
The corresponding fluctuation cut removes fractions
$5.47\times10^{-6}$ and $7.81\times10^{-6}$ of the original
prior volume for the broad and sub-Planckian decay-constant
ranges, respectively.
For $A_0$, extending the local likelihood profile to initial
displacements four orders of magnitude below the benchmark
threshold yields no further improvement in $\chi^2$ along
the explored branch, within the numerical resolution
(Sec.~\ref{subsec:aiccomparison}).

We define $x_f\equiv\log_{10}(f/\mpl)$,
$x_{f,\min}=\log_{10}(0.005)$, and $x_{f,\max}=1$.
The normalized joint density of $x_f$ and $\theta_i$ is
\begin{equation}
 \pi_Q(x_f,\theta_i)=
 \frac{\mathbf 1_{[x_{f,\min},x_{f,\max}]}(x_f)
       \mathbf 1_{[0,\,\pi-\delta_{\min}(x_f)]}(\theta_i)}
 {(x_{f,\max}-x_{f,\min})\pi q_{A_-}},
 \label{eq:quantum_joint_prior}
\end{equation}
where $\mathbf 1$ denotes an indicator and
$q_{A_-}=P_{\rm box}(Q)$ is the fraction of the original
joint box satisfying the quantum condition. Writing
$F_{\min,\max}=f_{\min,\max}/\mpl$, and noting that
$\delta_{\min}<\pi$ throughout the adopted range, this fraction is
\begin{align}
 q_{A_-}&=1-
 \frac{H_{\rm inf}/\mpl}
 {2\pi^2\ln(F_{\max}/F_{\min})}
 \left(\frac{1}{F_{\min}}-\frac{1}{F_{\max}}\right)
 \nonumber\\
 &=0.999994528342,
 \label{eq:quantum_volume}
\end{align}
using $\mpl=2.435\times10^{18}\,{\rm GeV}$.
For example, $\delta_{\min}=1.3072\times10^{-4}$ at
$f=0.005\mpl$ and $6.5361\times10^{-8}$ at $f=10\mpl$.

The cut correlates the prior for $f$ with its allowed angular
width. We implement Eq.~\eqref{eq:quantum_joint_prior} exactly
using independent uniform coordinates $u_f,u_\delta\in[0,1]$:
$x_f$ is obtained by inverting the cumulative density proportional
to $\pi-\delta_{\min}(x_f)$, and
$\delta_i=\delta_{\min}(x_f)+u_\delta[\pi-\delta_{\min}(x_f)]$.
This transformation preserves the original joint log-$f$,
uniform-angle measure conditioned on $Q$. In particular, sampling
$\log_{10}\delta_i$ uniformly would define a different prior.
The full domain is sampled in one PolyChord run, without
partitioning the initial angle into separate evidence runs.

For the spatially flat $\Lambda$CDM model, denoted by $L$,
we sample $(H_0,\omega_b,\omega_c,\sigma_8)$ with a uniform
prior $H_0\in[50,85]\,{\rm km\,s^{-1}\,Mpc^{-1}}$ and
the same priors on $\omega_b$, $\omega_c$, and $\sigma_8$.
In the axion models $H_0$ is determined by shooting the
background to satisfy present-day flatness; the same $H_0$
interval is applied as an acceptance criterion. For $L$,
spatial flatness determines the cosmological-constant density
for each sampled $H_0$. The quantum condition acts only on
the axion sector and leaves the $L$ reference prior unchanged.

\begin{table}[t]
\caption{Independent uniform priors for $A_-$ in the displayed
coordinates before conditioning on Eq.~\eqref{eq:quantum_cut}.
The normalized prior for each axion case is its underlying
joint prior conditioned on $Q$.}
\label{tab:axion_priors}
\begin{ruledtabular}
\begin{tabular}{lc}
Parameter & Underlying prior \\
\hline
$\omega_b$ & $[0.020,0.024]$ \\
$\omega_c$ & $[0.08,0.16]$ \\
$\log_{10}(\mphi/{\rm eV})$ & $[-34,-30]$ \\
$\log_{10}(f/\mpl)$ & $[\log_{10}(0.005),1]$ \\
$\OmL$ & $[0,3]$ \\
$\theta_i$ & $[0,\pi]$ \\
$\sigma_8$ & $[0.5,1.1]$ \\
\end{tabular}
\end{ruledtabular}
\end{table}

We restrict the analysis to solutions with $H>0$ and
$\rho_{\rm DE}>0$ throughout the evolution from initialization
to the present. With $\dot\phi(t_i)=0$, positivity of the initial
DE density requires $U(\phi_i)>\Lambda^{(-)}$, which is more
restrictive than Eq.~\eqref{eq:hilltopsupport}.
Along a regular expanding solution,
$\dot\rho_{\rm DE}=-3H\dot\phi^2\leq0$, so a positive DE density
today also guarantees positivity at earlier times.

For the algebraic reconstruction of $H^2$ in
Eq.~\eqref{eq:EN_alg}, we additionally require
$1-\Omega_{\rm kin}>0$, where
\begin{equation}
 \Omega_{\rm kin}\equiv\frac{\dot\phi^2}{6\mpl^2H^2}
 =\frac{\phi_N^2}{6\mpl^2}.
 \label{eq:omega_kin}
\end{equation}
For $H^2>0$, this is equivalent to $\rho_m+\rho_r+V>0$.
It avoids a vanishing denominator in Eq.~\eqref{eq:EN_alg}
and selects the domain connected to the frozen initial state
$\Omega_{\rm kin}(t_i)=0$. It is a numerical branch restriction,
not a no-ghost condition; excluded trajectories can be physical.

These branch and numerical criteria are selection tests
in addition to the sampled top-hat parameter bounds;
they are not observational likelihoods. Their treatment
in the evidence calculation is explained in
Sec.~\ref{subsec:polychord}.
The adopted parameterizations induce different prior
measures, particularly for $H_0$, which is derived in
the axion models but directly sampled in $\Lambda$CDM.
Moreover, since $\Lambda^{(-)}=3\mpl^2H_0^2\OmL$,
a uniform prior on $\OmL$ does not generally correspond
to a uniform prior on the dimensional vacuum-energy
density $\Lambda^{(-)}$ when $H_0$ varies.

\subsection{Background solver and flatness condition}
\label{subsec:numericalsolver}

For each sampled axion parameter set, 
we determine $H_0$ by imposing the present-day
Friedmann constraint. We integrate the background
equations \eqref{eq:kg_N}--\eqref{eq:EN_alg} from
$z_i=10^7$ to today (redshift $z=0$), with the initial conditions
\eqref{eq:initial_conditions}. The radiation density
is fixed by $T_{\rm CMB}=2.7255$ K and
$N_{\rm eff}=3.046$, assuming massless neutrinos.

For each trial value of $H_0$, we calculate the density
parameters $\Omega_{m0}=(\omega_b+\omega_c)/h^2$
and $\Omega_{r0}$ from the fixed physical matter and
radiation densities. We also update the dimensional
vacuum-energy density according to
$\Lambda^{(-)}=3\mpl^2H_0^2\OmL$, keeping the sampled
$\OmL$ fixed. We then vary $H_0$ within
\begin{equation}
 50~{\rm km\,s^{-1}\,Mpc^{-1}}<H_0
 <85~{\rm km\,s^{-1}\,Mpc^{-1}}\,,
 \label{eq:H0range}
\end{equation}
until the integrated background satisfies
\begin{equation}
 E^2(0)-1=0.
 \label{eq:H0shoot}
\end{equation}
We evaluate $E(0)$ using the trial $H_0$. At a shooting
root, the expansion rate obtained by evolving the background
to the present agrees with that same $H_0$.

We define the time-dependent displacement from the hilltop as
$\delta(N)\equiv\pi-\phi(N)/f$.
To resolve small positive displacements without subtracting
nearly equal angles, we evolve $y_\delta=\ln\delta$ and
$\eta=\dd\ln\delta/\dd N$ while $\delta<0.03$.
The Klein--Gordon equation then gives
\begin{align}
 y_{\delta,N}&=\eta,\nonumber\\
 \eta_N&=\frac{m_\phi^2}{H^2}\frac{\sin\delta}{\delta}
 -\left(3+\frac{H_N}{H}\right)\eta-\eta^2.
 \label{eq:log_hilltop_solver}
\end{align}
We evaluate $\sin\delta/\delta$ stably and switch at
$\delta=0.03$ to Eqs.~\eqref{eq:kg_N}--\eqref{eq:EN_alg},
matching the field and its first derivative. Initial points
with $\delta_i\geq0.03$ use the original variables; otherwise
the frozen condition gives $\eta(N_i)=0$.
The shooting starts and root selection depend only on the
current parameters, not on earlier likelihood evaluations.

Claude Code and ChatGPT assisted with code inspection and selected
numerical checks; tool details and author oversight are given in
the Acknowledgments.

The numerical integration employs DOP853 with a relative
tolerance of $2\times10^{-9}$, a maximum e-folding step
$\Delta N=0.04$, and 6000 stored points.
We compute $D_M$ by cumulative trapezoidal integration of
$c/H(z)$ on the stored redshift grid and use linear interpolation
for the distances and $H(z)$ required by the low-redshift likelihoods.
The shooting integrations are performed with a relative
tolerance of $8\times10^{-9}$.
We accept a shooting root satisfying
$|E^2(0)-1|\leq2\times10^{-7}$ and require
$|E^2(0)-1|<10^{-6}$ after the full integration.
We also verify that $H>0$, $\rho_{\rm DE}>0$,
and $1-\Omega_{\rm kin}>0$ throughout the integration,
and that the Friedmann constraint is satisfied.
We reject trials if the integration or shooting fails, the
results are nonfinite, or the numerical checks are not satisfied.
Such numerical rejection does not necessarily imply a physical
instability of the cosmological solution.
As noted in Sec.~\ref{sec:phenomenology}, this samples $f$
and shoots for $H_0$, reversing the illustrative normalization.

\subsection{SN, BAO, and compressed-CMB likelihoods}
\label{subsec:likelihoods}

Assuming statistical independence among the data sets,
we construct the joint likelihood as
\begin{equation}
 {\cal L}_{\rm tot}={\cal L}_{\rm SN}{\cal L}_{\rm BAO}
 {\cal L}_{\rm CMB}{\cal L}_{\rm RSD},
 \label{eq:likelihood_total}
\end{equation}
with the combined observations denoted by $\boldsymbol d$.
For a data sector $\ell\in\{{\rm SN},{\rm BAO},{\rm CMB},{\rm RSD}\}$,
we define $\chi_\ell^2\equiv-2\ln{\cal L}_\ell$ in the
common likelihood convention adopted here.
We use the same likelihood normalization and additive-constant
convention for all models, so common constants cancel in
Bayes factors and AIC differences.

For the DES Year 5 SN data, we employ the
\texttt{sn.desy5} likelihood \cite{DES:2024sn}.
For a supernova at redshift $z$, the distance modulus is
$\mu_{\rm th}(z)=5\log_{10}[D_L(z)/{\rm Mpc}]+25$,
up to a common offset associated with the SN absolute
magnitude. At the data level, \texttt{sn.desy5} evaluates
$D_A$ at the Hubble-diagram redshift and retains the distinct
heliocentric redshift in the luminosity prefactor.
We denote the observed distance moduli by
$\boldsymbol\mu_{\rm obs}$ and the predictions including this
redshift prescription by $\boldsymbol\mu_{\rm th}$.
Defining
$\boldsymbol\Delta\equiv
\boldsymbol\mu_{\rm obs}-\boldsymbol\mu_{\rm th}$,
we analytically marginalize over the common offset
to obtain
\begin{align}
 \chi^2_{\rm SN}
 &=\boldsymbol\Delta^{\rm T}
 \widetilde{\boldsymbol C}_{\rm SN}^{-1}
 \boldsymbol\Delta,
 \label{eq:snchi2}\\
 \widetilde{\boldsymbol C}_{\rm SN}^{-1}
 &=\boldsymbol C_{\rm SN}^{-1}
 -\frac{\boldsymbol C_{\rm SN}^{-1}\boldsymbol u
 \boldsymbol u^{\rm T}\boldsymbol C_{\rm SN}^{-1}}
 {\boldsymbol u^{\rm T}\boldsymbol C_{\rm SN}^{-1}
 \boldsymbol u},
 \label{eq:snmarg}
\end{align}
where $\boldsymbol C_{\rm SN}$ is the covariance and
$\boldsymbol u=(1,\ldots,1)^{\rm T}$ has one entry per SN.
This projects out the common offset; for fixed covariance and
the same flat offset measure, marginalization adds only a
common parameter-independent constant.

For the BAO data, we adopt the public
\texttt{bao.desi\_dr2} likelihood \cite{DESI:2025bao},
without including DESI full-shape clustering or
DESI RSD measurements.

For the CMB likelihood, we adopt the three-parameter
Gaussian compression presented in Appendix A of
Ref.~\cite{DESI:2025bao}, based on Ref.~\cite{Lemos:2023xhs}.
The theoretical prediction is specified by the vector
$\boldsymbol d_{\rm CMB}=(\theta_\star,\omega_b,\omega_{bc})$
defined in Eq.~\eqref{eq:cmbvector}. The corresponding
mean vector and covariance matrix are
\begin{equation}
 \overline{\boldsymbol d}_{\rm CMB}
 =(0.01041,0.02223,0.14208),
 \label{eq:cmbmean}
\end{equation}
and
\begin{equation}
 \boldsymbol C_{\rm CMB}=10^{-9}
 \begin{pmatrix}
 0.006621 & 0.12444 & -1.1929\\
 0.12444 & 21.344 & -94.001\\
 -1.1929 & -94.001 & 1488.4
 \end{pmatrix}.
 \label{eq:cmbcov}
\end{equation}
The CMB chi-square is then given by
\begin{equation}
 \chi^2_{\rm CMB}
 =(\boldsymbol d_{\rm CMB}
 -\overline{\boldsymbol d}_{\rm CMB})^{\rm T}
 \boldsymbol C_{\rm CMB}^{-1}
 (\boldsymbol d_{\rm CMB}
 -\overline{\boldsymbol d}_{\rm CMB}).
 \label{eq:cmbchi2}
\end{equation}
This compressed likelihood retains constraints on the
acoustic angular scale and physical matter densities,
but does not incorporate the full CMB anisotropy spectra
or their dependence on axion perturbations.

We evaluate the photon-decoupling and baryon-drag redshifts,
$z_\star$ and $z_{\rm drag}$, and their comoving sound horizons,
$r_\star$ and $r_{\rm drag}$, using quadratic fitting functions
calibrated against CAMB for reference flat $\Lambda$CDM
cosmologies. We retain the fitted redshifts and account
approximately for the modified expansion history of the axion
model by rescaling the fitted sound horizons according to
\begin{align}
 r_\beta&=r_\beta^{\rm fit}(\omega_b,\omega_m)
 \frac{I_\beta[H_{\rm axion}]}{I_\beta[H_{\rm ref}]},
 \label{eq:soundcorr}\\
 I_\beta[H]&\equiv\int_{z_\beta}^{\infty}
 \frac{c_s(z)}{H(z)}\,\dd z,
 \label{eq:soundint}
\end{align}
where $\beta=\star,{\rm drag}$ labels the two epochs and $r_\beta^{\rm fit}$
denotes the CAMB-calibrated sound-horizon fit.
Here $c_s(z)$ is the photon--baryon sound speed.
The expansion histories $H_{\rm axion}$ and
$H_{\rm ref}$ correspond to the axion model and
a reference flat $\Lambda$CDM model with identical
values of $H_0$, $\Omega_{m0}$, and $\Omega_{r0}$.
For the CMB acoustic scale, we evaluate $D_M(z_\star)$
by shape-preserving piecewise cubic interpolation of the tabulated
$D_M(z)$.
The same fitted lower limit $z_\beta$ and photon--baryon sound
speed are used in both integrals. Numerically, we use the
computed axion history up to $z_i=10^7$ and the same reference
tail for both integrals above this redshift.
This prescription modifies the background sound-horizon
integral while retaining the reference recombination calibration.
The sound-horizon fits are evaluated only within their
CAMB calibration domain,
\begin{equation}
 0.0214\leq\omega_b\leq0.0234,\qquad
 0.13\leq\omega_m\leq0.15.
 \label{eq:earlyfitdomain}
\end{equation}
The CMB acoustic scale uses $r_\star$, and the BAO distances
use $r_{\rm drag}$. Both sound horizons come from the same
calibrated fits. We therefore apply Eq.~\eqref{eq:earlyfitdomain}
to every model in the joint likelihood. The underlying priors
in Table~\ref{tab:axion_priors} are independent in $\omega_b$
and $\omega_c$, whereas the second calibration bound acts on
their sum $\omega_m$. Rejecting points outside the calibration
domain prevents extrapolation of the fits and changes the
accepted prior fraction in Sec.~\ref{subsec:polychord}.

To keep the early expansion history close to that assumed in
the compressed CMB likelihood, we also require
\begin{equation}
 0<\Omega_{\rm DE}(z_\star)<10^{-4},\qquad
 0<\Omega_{\rm DE}(z_{\rm drag})<10^{-4}.
 \label{eq:earlydecut}
\end{equation}
These bounds on the DE fraction at recombination and baryon drag
are likewise restrictions on the implemented prior support, not additional CMB likelihoods,
and do not by themselves establish consistency with the
full CMB anisotropy spectra.

\subsection{Growth integration and corrected RSD likelihood}
\label{subsec:rsdlikelihood}

To construct the RSD likelihood, we adopt the Gold-2017
compilation of 18 measurements over the redshift range
$0.02\leq z\leq1.40$ \cite{Nesseris:2017vor}.
We compute the theoretical growth observable
$f_g\sigma_8(z)$ by integrating Eq.~\eqref{eq:growth_approx}
from $N_g=-6$ to $N=0$, imposing the approximate
matter-era growing-mode initial conditions
\begin{equation}
 \delta_m(N_g)=e^{N_g},\qquad
 \delta_m'(N_g)=e^{N_g}.
 \label{eq:growthinitial}
\end{equation}
The numerical integration employs DOP853 with relative
and absolute tolerances of $2\times10^{-7}$ and
$10^{-10}$, respectively, a maximum step
$\Delta N=0.03$, and 500 interpolation points.
For each RSD redshift $z$,
we linearly interpolate $\delta_m'$ in $N$ to $N=-\ln(1+z)$
and use Eq.~\eqref{eq:fs8} to obtain $f_g(z)\sigma_8(z)$.
At $N_g=-6$, radiation is not entirely negligible:
for $\omega_m\simeq0.140$, we have
$\rho_r/\rho_m\simeq0.12$. Hence,
Eq.~\eqref{eq:growthinitial} approximates the
growing-mode solution rather than providing its
exact form in the radiation--matter transition.
At the best stored likelihood point and the largest-weight
sample of the $A_0$ production run, we repeated the growth
integration with $N_g=-8$ and $-10$, compared with the
baseline $N_g=-6$. With the 500-point interpolation used
in the RSD likelihood, the changes in $f_g\sigma_8(z=0.38)$
and $\chi^2_{\rm RSD}$ were below $1.2\times10^{-6}$ and
$1.7\times10^{-3}$, respectively. Direct evaluation of the
DOP853 solution reduced these differences to below
$3.2\times10^{-8}$ and $6.0\times10^{-6}$.
For these tested points, interpolation accounts for most of
the numerical variation; the normalized late-time growth
is insensitive to the starting time at this accuracy.

To account approximately for the fiducial cosmologies
adopted in the RSD measurements, we introduce the
Alcock--Paczynski correction factor
\begin{equation}
 \mathcal R_j\equiv\frac{H(z_j)D_A(z_j)}
 {H_j^{\rm fid}(z_j)D_{A,j}^{\rm fid}(z_j)},
 \qquad
 m_j=\mathcal R_j[f_g\sigma_8]_{\rm th}(z_j),
 \label{eq:rsdgeometry}
\end{equation}
where $j$ labels an RSD measurement and
$H_j^{\rm fid}$ and $D_{A,j}^{\rm fid}$ are
evaluated in the fiducial cosmology of the corresponding
survey. The dependence on the fiducial $H_0$ cancels
from their product, and radiation is neglected when
evaluating these low-redshift fiducial distances.
The RSD chi-square is then given by
\begin{equation}
 \chi^2_{\rm RSD}=
 (\boldsymbol d_{\rm RSD}-\boldsymbol m_{\rm RSD})^{\rm T}
 \boldsymbol C_{\rm RSD}^{-1}
 (\boldsymbol d_{\rm RSD}-\boldsymbol m_{\rm RSD}),
 \label{eq:rsdlikelihood}
\end{equation}
where $\boldsymbol d_{\rm RSD}$ and $\boldsymbol m_{\rm RSD}$ denote
the observed and predicted 18-component vectors,
respectively, and $\boldsymbol C_{\rm RSD}$ is
their covariance matrix. For the three correlated
WiggleZ measurements at $z=(0.44,0.60,0.73)$,
we adopt
\begin{equation}
 \boldsymbol C_{\rm WiggleZ}=10^{-3}
 \begin{pmatrix}
 6.400 & 2.570 & 0\\
 2.570 & 3.969 & 2.540\\
 0 & 2.540 & 5.184
 \end{pmatrix},
 \label{eq:wigglezcov}
\end{equation}
while neglecting cross-survey correlations, as in
the original compilation. The correction factor
$\mathcal R_j$ provides an approximate treatment of the
fiducial-geometry dependence rather than a
reanalysis of the survey multipoles.
We apply the same RSD likelihood prescription
to all models considered in the statistical analysis.

\subsection{PolyChord evidence and prior normalization}
\label{subsec:polychord}

PolyChord maintains $n_{\rm live}$ active parameter points and
successively replaces the lowest-likelihood point by one above
an increasing likelihood threshold. For the completed
seven-dimensional quantum-prior $A_-$ and $A_-^{\rm sub}$ runs
and the six-dimensional $A_0$ run, we use
\begin{align}
 n_{\rm live}&=800,\qquad n_{\rm prior}=2000,\nonumber\\
 n_{\rm repeat}&=28,\qquad {\rm precision}=0.01.
 \label{eq:polychordsettings}
\end{align}
Here $n_{\rm prior}$ counts accepted initialization points.
The setting $n_{\rm repeat}$ (\texttt{num\_repeats}) specifies the
number of one-dimensional slice-sampling steps used to generate a
replacement live point under the current likelihood constraint.
Each step can require several likelihood evaluations.
Sampling stops when $\widehat Z_{\rm live}/Z_{\rm acc}<0.01$, where
$\widehat Z_{\rm live}$ estimates the evidence remaining in the live
points and $Z_{\rm acc}$ is the evidence accumulated from discarded
points. This stopping rule controls the estimated residual contribution.
The statistical uncertainty on the final log evidence is quoted separately.

Each axion production run samples the full angular domain in
one calculation, without posterior boosting. Posterior summaries
are computed from the global weighted output without removing
a burn-in segment.
The unchanged four-dimensional $L$ reference used
$n_{\rm live}=800$, $n_{\rm prior}=1000$,
$n_{\rm repeat}=8$, and the same precision setting.

For each model $X$, we denote the underlying rectangular
prior domain by $\mathcal B_X$ and its volume by
$\mathcal V_X=\prod_i(\vartheta_{i,\max}-\vartheta_{i,\min})$.
Here $\vartheta_i$ denotes one coordinate of the parameter vector
$\boldsymbol\vartheta$. For axions, the volume element uses the
logarithms of $m_\phi$ and $f$ and the initial angle $\theta_i$.
We denote the set satisfying Eq.~\eqref{eq:quantum_cut} by $Q_X$
and define $\mathcal C_X=\mathcal B_X\cap Q_X$ and
$q_X=\operatorname{Vol}(\mathcal C_X)/\mathcal V_X$. For $L$
the entire box is retained, so $\mathcal C_L=\mathcal B_L$
and $q_L=1$. The normalized prior before numerical acceptance is
\begin{equation}
 \pi_{Q,X}(\boldsymbol\vartheta)
 =\frac{\mathbf 1_{\mathcal C_X}(\boldsymbol\vartheta)}
 {q_X\mathcal V_X}.
 \label{eq:normalized_quantum_prior}
\end{equation}
The indicator $\mathbf 1_{\mathcal C_X}$ equals one inside
$\mathcal C_X$ and zero elsewhere. 
The sampling procedure described in Sec.~\ref{subsec:priors}
generates samples directly from this normalized prior.
Its evidence therefore already includes the normalization
by $q_X$.

We denote the accepted subset of $\mathcal C_X$ by $\mathcal A_X$.
All models obey the sound-horizon calibration bounds
\eqref{eq:earlyfitdomain}. For axions we also require an expanding
background with $\rho_{\rm DE}>0$ and $1-\Omega_{\rm kin}>0$,
the DE-fraction bounds in Eq.~\eqref{eq:earlydecut}, and a successful flatness
solve to the tolerances in Sec.~\ref{subsec:numericalsolver}.
Consequently, the accepted volume depends on the chosen branch,
the range of the calibration, and the numerical solver.

We define the accepted fraction relative to the already normalized
quantum prior by
\begin{equation}
 p_X\equiv P_{Q,X}(\mathcal A_X)
 =\frac{\operatorname{Vol}(\mathcal A_X)}{q_X\mathcal V_X}.
 \label{eq:accepted_fraction_quantum}
\end{equation}
Here $P_{Q,X}$ denotes probability under the normalized prior
$\pi_{Q,X}$. Convention I uses $\pi_{Q,X}$ and assigns zero
likelihood to points outside the accepted region $\mathcal A_X$.
Convention II conditions $\pi_{Q,X}$ on $\mathcal A_X$, dividing
the prior density in that region by $p_X$.

With $D=\dim\mathcal B_X$ and likelihood density
$\mathcal L_X=P(\boldsymbol d\mid\boldsymbol\vartheta,X)$,
the evidences are

\begin{align}
 Z_X^{\rm I}
 &=\frac{1}{q_X\mathcal V_X}\int_{\mathcal A_X}{\cal L}_X
 \,\dd^D\boldsymbol\vartheta,\\
 Z_X^{\rm II}
 &=\frac{1}{p_Xq_X\mathcal V_X}\int_{\mathcal A_X}{\cal L}_X
 \,\dd^D\boldsymbol\vartheta.
 \label{eq:evidence_conventions_quantum}
\end{align}
Thus $Z_X^{\rm I}=p_XZ_X^{\rm II}$. Conditioning the prior on
$\mathcal A_X$ divides both its density and the evidence by
$p_X$. These factors cancel in the normalized posterior, even
though they change the evidence and hence the Bayes factor.
The native PolyChord output uses Convention II.

The accepted fraction is estimated during initialization as
\begin{equation}
 f_{{\rm valid},X}=\frac{n_{\rm prior}}{n_{{\rm draw},X}},
 \label{eq:fvalid}
\end{equation}
where $n_{{\rm draw},X}$ counts all draws from $\pi_{Q,X}$,
including those rejected by the fixed acceptance tests.
For all three completed axion runs, the quantum cut is built into these
draws; $f_{{\rm valid},X}$ is therefore distinct from the
geometric fraction $q_X$ retained from the original box.
The numerical estimates of the log evidences are
\begin{align}
 \ln Z_X^{\rm I}
 &=\ln Z_{{\rm raw},X}+\ln f_{{\rm valid},X},
 \label{eq:evidenceI}\\
 \ln Z_X^{\rm II}&=\ln Z_{{\rm raw},X}.
 \label{eq:evidenceII}
\end{align}
Native PolyChord statistics report the evidence for
$\pi_{Q,X}/p_X$ on $\mathcal A_X$; processed Cobaya output
already includes $\ln f_{{\rm valid},X}$ and uses Convention I.
For deterministic acceptance, $f_{{\rm valid},X}\to p_X$ as
$n_{\rm prior}$ increases, counting numerical failures as rejections.

The common sound-horizon calibration domain defined by
Eq.~\eqref{eq:earlyfitdomain} occupies
\begin{equation}
 f_{\rm cal}=
 \frac{0.0234-0.0214}{0.024-0.020}
 \frac{0.15-0.13}{0.16-0.08}=0.125
 \label{eq:fcal}
\end{equation}
of the $(\omega_b,\omega_c)$ prior. This result follows from
the unit Jacobian of the transformation from
$(\omega_b,\omega_c)$ to $(\omega_b,\omega_m)$; the calibration
rectangle lies wholly inside the transformed prior domain.
The quantum condition depends only on $f$ and $\theta_i$,
so this common factor is
unchanged. It is included in $f_{{\rm valid},X}$ and cancels
in matched Bayes factors. Other support restrictions can differ
between models.

For independent prior draws stopped after $n_{\rm prior}$
acceptances, the log of the acceptance fraction in
Eq.~\eqref{eq:fvalid} has leading counting variance
\begin{equation}
 \sigma^2(\ln f_{{\rm valid},X})\simeq
 \frac{1-f_{{\rm valid},X}}{n_{\rm prior}}.
 \label{eq:counting_variance_quantum}
\end{equation}
Assuming independent counting and nested-sampling errors, we add
this term to the PolyChord log-evidence variance for Convention I.
The quoted errors do not quantify sensitivity to alternative
priors or likelihood approximations, or errors caused by
solver-dependent exclusions or posterior modes missed by a finite run.

\section{Observational constraints}
\label{sec:results}

\subsection{Negative-vacuum posterior constraints}
\label{subsec:posterior}

\begin{figure*}[t]
 \centering
 \includegraphics[width=0.92\textwidth]{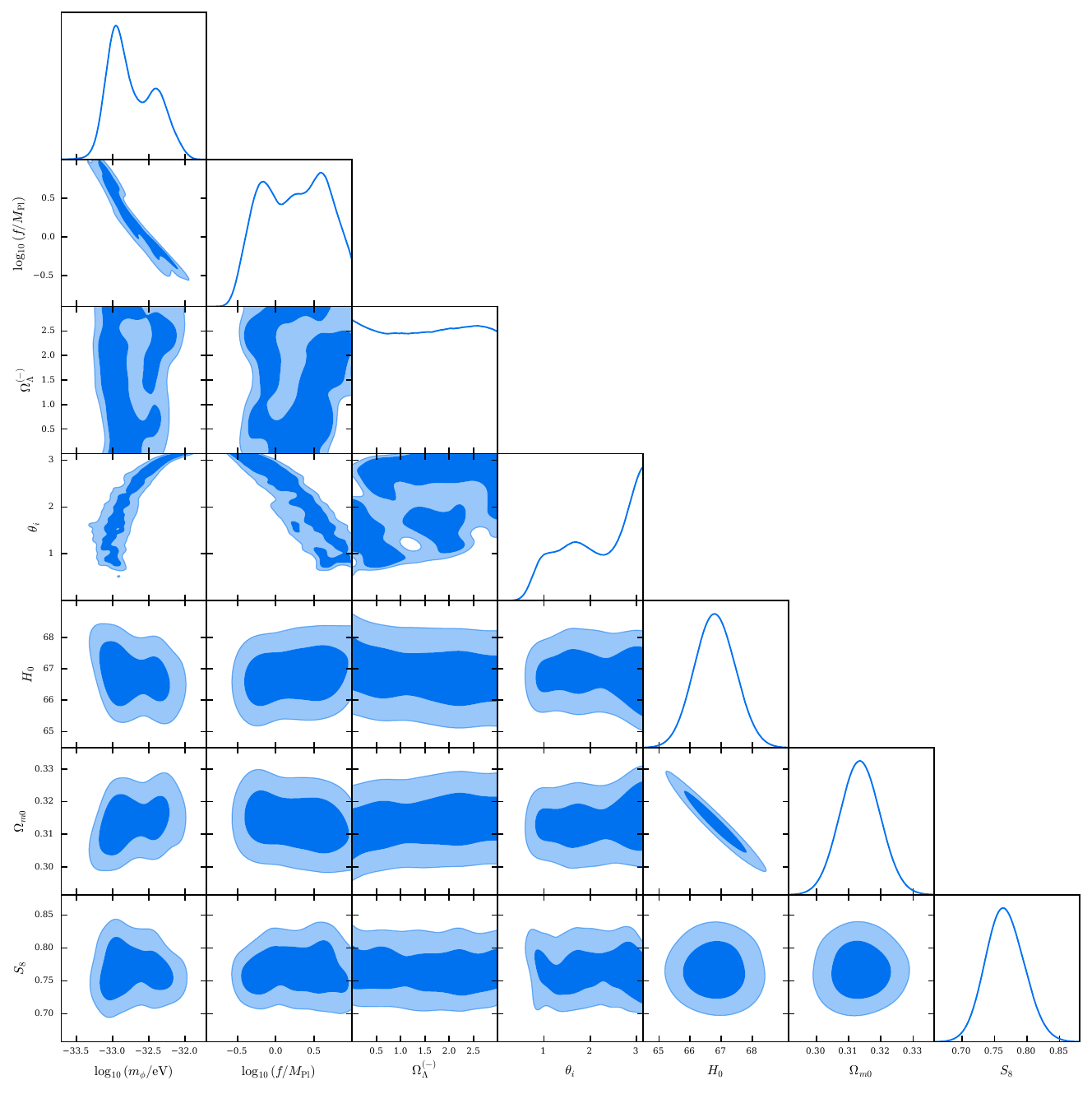}
 \caption{Marginalized posterior constraints for the
 quantum-prior $A_-$ analysis, using DES Year 5 SN,
 DESI DR2 BAO, compressed CMB, and corrected Gold-2017 RSD.
 The underlying range is $0.005\leq f/\mpl\leq10$,
 conditioned on Eq.~\eqref{eq:quantum_cut}.
 The seven displayed quantities are
 $\{\log_{10}(m_\phi/{\rm eV}),\log_{10}(f/\mpl),\OmL,
 \theta_i,H_0,\Omega_{m0},S_8\}$.
 The inner and outer filled density contours enclose
 approximately 68\% and 95\% of each two-dimensional
 marginalized posterior. They are not one-dimensional
 equal-tail intervals. Diagonal panels show marginalized
 one-dimensional densities.}
 \label{fig:triangle}
\end{figure*}

Figure~\ref{fig:triangle} and Table~\ref{tab:negativeLambdaPosterior}
show the broad $A_-$ posterior for the priors of
Sec.~\ref{subsec:priors}. The run has 16804 dead points,
16760 posterior points in the final statistics, and
$2.658835\times10^6$ likelihood evaluations; the global
weighted export contains 14612 rows. We scale weights $W_i$
to $\max_i W_i=1$ and normalize by $\sum_iW_i$ for summaries.
The weight-only effective size
$(\sum_iW_i)^2/\sum_iW_i^2\simeq5190$ measures weight
dispersion, not the number of independent samples. The seven
recorded clusters are algorithmic groups, not established
physical branches or evidence of exhaustive mode recovery.

The well-constrained late-time parameters have means
$H_0=66.80\,{\rm km\,s^{-1}\,Mpc^{-1}}$,
$\Omega_{m0}=0.3136$, $\sigma_8=0.749$, and $S_8=0.766$.
The posterior standard deviations of $H_0$ and $S_8$ are
$0.62\,{\rm km\,s^{-1}\,Mpc^{-1}}$ and $0.029$,
respectively. The inferred Hubble constant is not shifted toward the higher
local-distance-ladder determinations. We do not include a
local-$H_0$ likelihood or quantify this discrepancy here.

The axion parameters remain correlated and non-Gaussian.
The mass marginal has multiple peaks, so its mean and central
interval alone do not describe the density shown in
Fig.~\ref{fig:triangle}.
The exact identity
\begin{align}
 \mphi^2f^2\sin^2\left(\frac{\phi_0}{2f}\right)
 &=\frac{3}{4}(1-w_{\phi,0})\Omega_{\phi,0}\mpl^2H_0^2
 \nonumber\\
 &=\frac{3}{4}\mpl^2H_0^2
 \left[(1-w_{{\rm DE},0})\Omega_{{\rm DE},0}\right.
 \nonumber\\
 &\quad\left.{}+2\OmL\right]
 \label{eq:mf_posterior_identity}
\end{align}
follows from Eqs.~\eqref{eq:potential} and
\eqref{eq:wphi_wde_relation}. For $\OmL\ne0$, the axion equation of state $w_\phi$
and the total DE equation of state $w_{\rm DE}$ generally differ.
Eq.~\eqref{eq:wphi_wde_relation} supplies the $2\OmL$ term
when the mass relation is written in terms of $w_{\rm DE}$. 
The weighted standard deviation of
$\log_{10}[(m_\phi/{\rm eV})(f/\mpl)]$ is $0.134$,
smaller than $0.315$ for $\log_{10}(m_\phi/{\rm eV})$ and
$0.406$ for $\log_{10}(f/\mpl)$.
At fixed $\phi_0/f$ and fixed background quantities on the
right-hand side,
Eq.~\eqref{eq:mf_posterior_identity} gives $m_\phi f=\mathrm{const.}$,
or $m_\phi\propto f^{-1}$ with logarithmic slope $-1$.
The field ratio and background quantities, including $\OmL$,
vary across the posterior, allowing the shallower weighted
slope $-0.75$ when we regress $\log_{10}(m_\phi/{\rm eV})$
on $\log_{10}(f/\mpl)$ with the posterior weights. Variation in the field angle and the background densities thus contributes
to the width of the mass--decay-constant correlation.

The marginal central 68\% intervals are
\begin{align}
 m_\phi&\in[0.91,4.39]\times10^{-33}\,{\rm eV},\\
 f/\mpl&\in[0.589,4.899].
 \label{eq:axion_quantum_intervals}
\end{align}
The equal-tail 95\% credible interval for $f/\mpl$ is
$[0.345,8.326]$. Its upper tail remains sensitive to the imposed
$f/\mpl\leq10$ boundary.
The fifth percentile of $f$ and the 95th percentile of
$m_\phi$ give the one-sided 95\% posterior limits
$f/\mpl>0.396$ and $m_\phi<6.56\times10^{-33}\,{\rm eV}$,
respectively.

The negative-vacuum parameter remains weakly constrained,
with a posterior mean of $1.501$ and a central 68\% credible
interval of $[0.473,2.492]$.
The weighted Pearson coefficients are $+0.24$ for $\OmL$ and
$w_{{\rm DE},0}$, and $+0.33$ for $\OmL$ and $\theta_i$.
These weak correlations include variations in $H_0$ and the
axion parameters, unlike the illustrative fixed-$H_0$ family
in Fig.~\ref{fig:fig3_oml}.
An equal-tail interval with a positive lower endpoint under
the continuous $\OmL\geq0$ prior does not establish a preference
over the separate model $A_0$. The matched evidence comparison
is given in Sec.~\ref{subsec:modelcomparison}.

\begin{table}[t]
\caption{Posterior means and central 68\% credible intervals
for the completed quantum-prior $A_-$ run with
$0.005\leq f/\mpl\leq10$. Intervals are the weighted
$[\mathcal Q_{0.16},\mathcal Q_{0.84}]$ endpoints, where
$\mathcal Q_p$ is the weighted $p$-quantile. The Hubble constant is in
${\rm km\,s^{-1}\,Mpc^{-1}}$; the other displayed variables
are dimensionless. The mean and interval midpoint need not
coincide for a non-Gaussian posterior.}
\label{tab:negativeLambdaPosterior}
\begin{ruledtabular}
\begin{tabular}{lcc}
Parameter & Mean & 68\% interval \\
\hline
$\log_{10}(m_\phi/{\rm eV})$ & $-32.730$ & $[-33.041,-32.358]$ \\
$\log_{10}(f/\mpl)$ & $0.241$ & $[-0.230,0.690]$ \\
$\OmL$ & $1.501$ & $[0.473,2.492]$ \\
$\theta_i$ & $2.105$ & $[1.209,2.948]$ \\
$H_0$ & $66.80$ & $[66.17,67.44]$ \\
$\Omega_{m0}$ & $0.3136$ & $[0.3076,0.3195]$ \\
$\sigma_8$ & $0.749$ & $[0.720,0.780]$ \\
$S_8$ & $0.766$ & $[0.738,0.796]$ \\
$w_{{\rm DE},0}$ & $-0.851$ & $[-0.918,-0.797]$ \\
\end{tabular}
\end{ruledtabular}
\end{table}

\subsection{Reconstructed equation of state}
\label{subsec:wde}

\begin{figure}[tbp]
 \centering
 \includegraphics[width=0.95\linewidth]{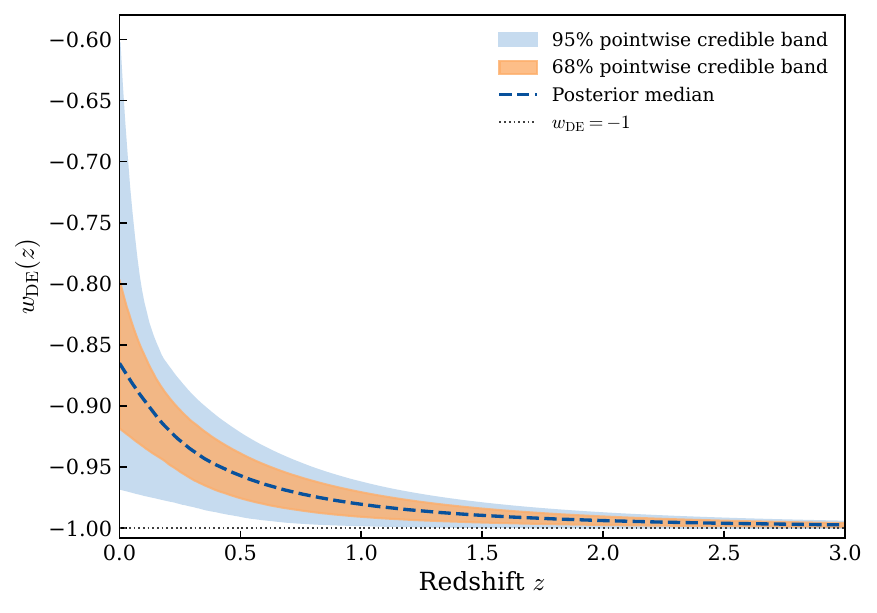}
 \caption{Reconstructed $w_{\rm DE}(z)$ for the full
 quantum-prior $A_-$ posterior. The dashed curve is the
 pointwise posterior median. The shaded regions show
 central pointwise 68\% and 95\% credible intervals,
 and the horizontal dotted line marks $w_{\rm DE}=-1$.
 The reconstruction uses 11394 weighted histories
 at 301 redshifts in $0\leq z\leq3$.}
 \label{fig:wdeposterior}
\end{figure}

Figure~\ref{fig:wdeposterior} reconstructs the total-DE
equation of state from the same full quantum-prior posterior.
We evolve 11394 weighted parameter points forward from
$z_i=10^7$ with the production background solver and evaluate
their histories at 301 redshifts in $0\leq z\leq3$.
We retain points with $W_i>10^{-12}$ in the normalization of
Sec.~\ref{subsec:posterior}. All retained histories integrate
successfully and reproduce the saved $w_{{\rm DE},0}$ to within
$1.25\times10^{-12}$. Quantiles are calculated with the
posterior weights at each redshift.

At $z=0$, the pointwise posterior median is
\begin{equation}
 w_{{\rm DE},0}^{\rm med}=-0.865,
 \label{eq:w0_quantum_median}
\end{equation}
with central 68\% and 95\% intervals
$[-0.918,-0.797]$ and $[-0.968,-0.603]$, respectively.
The mean in Table~\ref{tab:negativeLambdaPosterior} differs
from the median because the distribution has a tail toward
larger $w_{{\rm DE},0}$.
At $z=0.5$ and $z=1$, the medians are $-0.957$ and $-0.981$,
respectively, with corresponding central 68\% intervals
$[-0.976,-0.939]$ and $[-0.991,-0.971]$.
The $z=3$ median of $-0.9974$ is consistent with thawing
from an almost frozen field.

We compute the median and central 68\% and 95\% intervals
separately at each redshift. Different solutions can determine
the median at different redshifts, so the median curve need
not trace any single physical trajectory. The quoted coverage
applies pointwise. All histories obey $w_{\rm DE}\geq-1$
on the selected positive-density branch [Eq.~\eqref{eq:nonnull}].

\subsection{Akaike information criterion comparison}
\label{subsec:aiccomparison}

The AIC \cite{Akaike:1974} balances goodness of fit against
model complexity by penalizing the number of fitted parameters:
\begin{equation}
 {\rm AIC}=-2\ln{\cal L}_{\max}+2N_{\rm par}
          =\chi^2_{\min}+2N_{\rm par}.
 \label{eq:aic_definition}
\end{equation}
Here $\mathcal L_{\max}$ is the likelihood maximized over the
model's allowed parameter domain, and
$\chi^2_{\min}=-2\ln\mathcal L_{\max}$ uses the common likelihood
normalization adopted for all models.
The quantity $N_{\rm par}$ is the number of independently fitted
cosmological parameters.
We define the difference between models $X$ and $Y$ as
$\Delta{\rm AIC}_{X/Y}\equiv{\rm AIC}_X-{\rm AIC}_Y$,
so negative values favor $X$ over $Y$ under this criterion.

Table~\ref{tab:aic} distinguishes the smallest $\chi^2$ among
saved PolyChord samples, $\chi^2_{\rm chain}$, from the endpoint
of an additional local minimization, $\chi^2_{\rm local}$.
The PolyChord stopping rule controls the remaining evidence and
does not specify the accuracy of a likelihood maximum. Local
searches can also miss lower values elsewhere. Every admissible
point nevertheless supplies an upper bound on $\chi^2_{\min}$,
including a point inherited from a nested model.

The reference model $L$ fits $(H_0,\omega_b,\omega_c,\sigma_8)$.
The two negative-vacuum models fit $(\omega_b,\omega_c,m_\phi,f,
\theta_i,\OmL,\sigma_8)$, with $\OmL$ free over its full prior
range including zero. Fixing $\OmL=0$ leaves six parameters in
$A_0$. In each axion model the present-day flatness condition
determines $H_0$ (Sec.~\ref{subsec:numericalsolver}). We fix
$H_{\rm inf}$ and use the same SN-offset marginalization for
every model. Reparametrizations used for sampling preserve these
parameter counts.

For $L$, $\chi^2_{\rm chain}=1675.6231$ and
$\chi^2_{L,{\rm local}}=1675.609881$.
The searches in Table~\ref{tab:aic} retain the production
likelihood and acceptance criteria, including the quantum cut;
the profile below relaxes only that cut. Reproduction checks
and search settings are given below and in
Appendix~\ref{app:local_searches}.

For $A_-$, independent re-evaluation of the point selected from
the initial searches gives $\chi^2=1660.752979$. A further local
refinement lowers this value by $9.56\times10^{-6}$ to

\begin{equation}
 \chi^2_{A_-,{\rm local}}=1660.752969.
 \label{eq:bestfit_local_refinement}
\end{equation}
The $A_-$ endpoint has $f/\mpl=0.0888293$,
$\delta_i=1.52204\times10^{-5}$, and
$\delta_{\min}=7.35809\times10^{-6}$, giving
$\delta_i/\delta_{\min}\simeq2.07$ and admissibility in
$A_-^{\rm sub}$. The lower $A_0$ point discussed below supplies
upper bounds for both negative-vacuum models.

\begin{table*}[tbp]
\caption{Likelihood-search results and AIC benchmarks.
$\chi^2_{\rm chain}$ is the smallest saved PolyChord value;
$\chi^2_{\rm local}$ is the endpoint of an additional local
minimization, including refinement. The $L$ and $A_0$ AICs are
numerical estimates. The bounds for $A_-$ and $A_-^{\rm sub}$
use the common admissible $\OmL=0$ point, independently of the listed local
endpoints. The final column subtracts the local-fit AIC estimate
$\widehat{\rm AIC}_L\equiv\chi^2_{L,{\rm local}}+8=1683.609881$.
The shared point lies near the quantum-cut boundary. The listed
endpoints are local fits.}
\label{tab:aic}
\begin{ruledtabular}
\begin{tabular}{lccccc}
Model & $N_{\rm par}$ & $\chi^2_{\rm chain}$ & $\chi^2_{\rm local}$ & ${\rm AIC}$ & ${\rm AIC}-\widehat{\rm AIC}_L$ \\
\hline
$L$ & 4 & $1675.6231$ & $1675.6099$ & $\simeq1683.6099$ & $\simeq0$ \\
$A_0$ & 6 & $1661.8594$ & $1660.7418$ & $\simeq1672.7418$ & $\simeq-10.8680$ \\
$A_-$ & 7 & $1661.4449$ & $1660.7530$ & $\leq1674.7418$ & $\leq-8.8680$ \\
$A_-^{\rm sub}$ & 7 & $1661.1954$ & $1660.7418$ & $\leq1674.7418$ & $\leq-8.8680$ \\
\end{tabular}
\end{ruledtabular}
\end{table*}

For $A_0$, additional local $\chi^2$ minimization starting from
saved chain samples yields $\chi^2_{A_0,{\rm local}}=1660.741841593$,
compared with the smallest saved-chain value of $1661.859436889$.
We refer to the solution attaining this value as the refined local
best fit, without claiming global optimality.
Using identical data and likelihood normalization, an independent
evaluation reproduces the refined value within $2\times10^{-8}$,
while a fresh four-parameter minimization reproduces the local
$L$ fit within $5\times10^{-7}$.
The contributions from SN, BAO, CMB, and RSD are
$(1635.189442,10.329536,2.000322,13.222541)$ for $A_0$ and
$(1649.6346,12.2278,1.0021,12.7453)$ for $L$, respectively.
The improvement relative to $L$,
$\Delta\chi^2\equiv\chi^2_{A_0,{\rm local}}
-\chi^2_{L,{\rm local}}\simeq-14.8680$,
comes predominantly from DES Year 5 SN ($-14.445$), with an additional
contribution from BAO ($-1.898$), partly offset by increases
in the CMB ($+0.998$) and RSD ($+0.477$) contributions.

This refined local best-fit $A_0$ solution has $f/\mpl=0.0630932$,
$m_\phi=2.38815\times10^{-32}\,{\rm eV}$,
$H_0=65.81556\,{\rm km\,s^{-1}\,Mpc^{-1}}$,
and $\sigma_8=0.752283$.
Its $\delta_i=1.0359665\times10^{-5}$ lies just above
the quantum-cut boundary $\delta_{\min}=1.0359502\times10^{-5}$,
with $\delta_i/\delta_{\min}=1.000016$.
This proximity alone does not establish whether relaxing
the quantum cut would further improve the fit.

We define the present-day deceleration parameter as
\begin{equation}
 q_0\equiv-\frac{\ddot a_0}{a_0H_0^2}
 =\frac{1+\Omega_{r0}+3w_{{\rm DE},0}\Omega_{{\rm DE},0}}{2}.
 \label{eq:profile_q0}
\end{equation}
Negative and positive values of $q_0$ indicate acceleration and
deceleration, respectively.
For this local best-fit $A_0$ solution, $m_\phi/H_0\simeq17$ and
$w_{{\rm DE},0}=+0.07588$, giving $q_0\simeq+0.577$.
The solution is therefore decelerating today, which the acceptance
criteria permit.
The $A_0$ AIC estimate uses this locally optimized point with the
smooth-DE growth likelihood of Sec.~\ref{subsec:observables}.
Posterior constraints and evidences are obtained from the production runs.

Since the refined local best-fit $A_0$ point has $f<\mpl$,
it is admissible in both negative-vacuum models at $\OmL=0$.
At this boundary, their background evolution and likelihood
coincide with those of $A_0$ for the same values of the six
shared parameters. We define the common reference value
\begin{equation}
 b_0\equiv\chi^2_{A_0,{\rm local}}=1660.741841593.
 \label{eq:b0_definition}
\end{equation}
An independent local minimization over all seven parameters
of $A_-^{\rm sub}$ yields $\chi^2=1660.741842$ at
$\OmL\simeq4.4\times10^{-8}$; the unrounded $\chi^2$ agrees
with $b_0$ to within $5\times10^{-8}$.
Convergence tests and a reevaluation with a fresh model instance
are documented in Appendix~\ref{app:local_searches}.
This endpoint is also admissible in $A_-$; no additional search
over the full $A_-$ domain initialized at $\OmL=0$ is reported here.

The parameter domain of $A_-^{\rm sub}$ is contained in that
of $A_-$ and includes the $A_0$ boundary point defining $b_0$.
The global minima therefore satisfy
\begin{equation}
 \chi^2_{A_-,\min}\leq\chi^2_{A_-^{\rm sub},\min}
 \leq b_0.
 \label{eq:subplanck_minimum_bound}
\end{equation}
The second inequality uses this particular $A_0$ point,
not the minimum over the full $A_0$ domain, which also
includes $f>\mpl$.
The local $A_-$ solution in Eq.~\eqref{eq:bestfit_local_refinement}
has a $\chi^2$ exceeding $b_0$ by approximately $0.01113$
and is therefore not globally optimal.
The local searches within the restricted domain likewise do not
rule out lower $\chi^2$ values elsewhere in that domain.
Accordingly, Table~\ref{tab:aic} lists the locally optimized
values but uses $b_0$ to obtain AIC upper bounds for both models.

The models $A_-$ and $A_-^{\rm sub}$ each have $N_{\rm par}=7$.
The parameter $\OmL$ is fitted even when its optimum lies at zero,
and restricting the allowed range of $f$ does not reduce the
parameter count. The common admissible $A_0$ point therefore gives
\begin{align}
 {\rm AIC}_X&\leq b_0+14=1674.741841593,\nonumber\\
 {\rm AIC}_X-\widehat{\rm AIC}_L&\lesssim-8.8680,
 \quad X\in\{A_-,A_-^{\rm sub}\},
 \label{eq:aic_result}
\end{align}
where
$\widehat{\rm AIC}_L\equiv\chi^2_{L,{\rm local}}+8=1683.609881$
is the AIC estimate from the locally optimized flat $\Lambda$CDM
($L$) fit. The second line is rounded and refers to this fixed
local reference; global optimality of the $L$ fit has not
been established.

\begin{figure}[t]
\centering
\includegraphics[width=\linewidth]{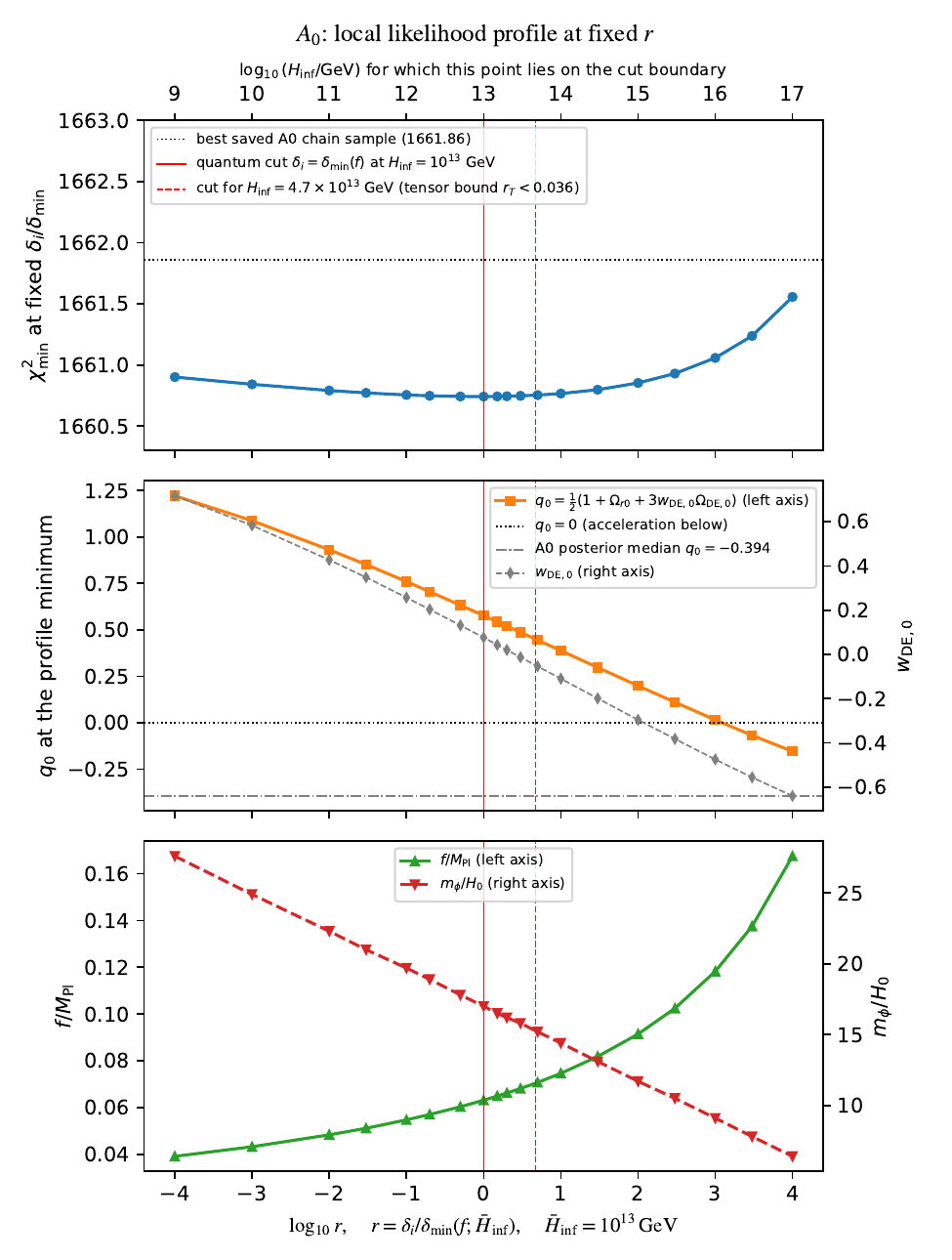}
\caption{Local $A_0$ profile at fixed
$r=\delta_i/\delta_{\min}(f;\bar H_{\rm inf})$ with
$\bar H_{\rm inf}=10^{13}\,{\rm GeV}$, minimizing the remaining
five parameters. The panels show the attained $\chi^2$,
the deceleration parameter $q_0$ (orange, left axis) with
$w_{{\rm DE},0}$ (grey, right axis), and $f/\mpl$ (green, left
axis) with $m_\phi/H_0$ (red, right axis). The upper axis maps a
point to the inflationary scale for which it lies on the cut
boundary; at that scale all points to its right are admissible.
The vertical lines mark the benchmark cut ($r=1$) and the cut at
the BK18 upper-limit scale ($r\simeq4.7$). The $q_0=0$ line
separates acceleration and deceleration, and the dash-dotted
horizontal line marks the $A_0$ posterior median.
Connecting lines guide the eye; global minima are not certified.}
\label{fig:a0profile}
\end{figure}

For the six-parameter $A_0$ model, the refined local best fit yields
\begin{align}
 \widehat{\rm AIC}_{A_0}&=b_0+12\simeq1672.741842,\nonumber\\
 \widehat{\Delta{\rm AIC}}_{A_0/L}&\simeq-10.8680.
 \label{eq:A0_aic_result}
\end{align}
The hats denote estimates based on the local fits for $A_0$ and $L$.
These give an AIC preference for the zero-vacuum axion model
$A_0$ over flat $\Lambda$CDM, with
$\widehat{\Delta{\rm AIC}}_{A_0/L}\simeq-10.87$.
To assess whether relaxing the quantum cut allows a further
improvement in the $A_0$ fit, we construct a local $\chi^2$ profile
using the displacement ratio
\begin{equation}
 r\equiv\frac{\delta_i}{\delta_{\min}(f;\bar H_{\rm inf})},
 \qquad \bar H_{\rm inf}=10^{13}\,{\rm GeV},
 \label{eq:profile_ratio}
\end{equation}
where $\delta_{\min}(f;\bar H_{\rm inf})=\bar H_{\rm inf}/(2\pi f)$.
At each of 19 fixed values of $r$ between $10^{-4}$ and $10^4$,
we minimize $\chi^2$ locally over the remaining five parameters.
Figure~\ref{fig:a0profile} shows the resulting profile,
$\chi^2_{\rm prof}(r)$.
The production prior imposes $r\geq1$.
The smallest $r$ among the 14\,557 saved weighted $A_0$ samples
is $3.52$, whereas the refined local best-fit $A_0$ point
has $r=1.000016$.
Thus, the saved samples do not resolve the likelihood near the
quantum-cut boundary. The profile examines both this region and
the smaller initial displacements at $r<1$ excluded from the
production prior.

The locally minimized $\chi^2_{\rm prof}(r)$ exceeds $b_0$
by $0.013$ at $r=0.1$ and $0.16$ at $r=10^{-4}$.
For $r>1$, the corresponding increases are $0.024$ at $r=10$
and $0.11$ at $r=100$.
Over $0.5\lesssim r\lesssim2$, the variation is at most $0.003$,
comparable to the numerical resolution of $0.001$--$0.003$.
Thus, extending the local search below the benchmark quantum cut
down to $r=10^{-4}$ reveals no improvement over $b_0$ along
the sampled branch within this resolution.
An independent hilltop solver reproduces the same qualitative
dependence on $r$ at six fixed ratios, with absolute differences
in $\chi^2$ of $0.001$--$0.014$ between the two implementations
(Appendix~\ref{app:local_searches}).

Changing $H_{\rm inf}$ shifts the admissible domain to
\begin{equation}
 r\geq\frac{H_{\rm inf}}{\bar H_{\rm inf}}.
 \label{eq:profile_cut_mapping}
\end{equation}
For a given $H_{\rm inf}$, the best fit along the explored branch
is obtained by minimizing $\chi^2_{\rm prof}(r)$ over the allowed
values of $r$.
Lowering $H_{\rm inf}$ keeps the refined local best-fit $A_0$
point at $r\simeq1$ admissible and allows the smaller displacements
tested above.
To test a stronger quantum cut, we choose a higher $H_{\rm inf}$
near the observational upper limit inferred from BK18.
BK18 constrains the primordial tensor-to-scalar ratio $r_T$ to
$r_T<0.036$ at 95\% confidence \cite{BICEPKeck:2021}.
Under the inflationary assumptions of Sec.~\ref{subsec:priors},
$H_*^2=(\pi^2/2)\mpl^2A_sr_T$ with $A_s\simeq2.1\times10^{-9}$
gives $H_*\lesssim4.7\times10^{13}\,{\rm GeV}$.
Taking $H_{\rm inf}\simeq H_*$ near this limit places the
quantum-cut boundary at
$r=H_{\rm inf}/\bar H_{\rm inf}\simeq4.7$, since
$\bar H_{\rm inf}=10^{13}\,{\rm GeV}$.
Thus, $r\simeq4.7$ tests the fit when the quantum cut is tightened
to the inflationary scale corresponding to the BK18 upper limit.
The locally optimized fit near this new boundary has
$\chi^2-b_0\simeq0.01$, with the same increase in the estimated AIC.
The dependence of the evidence and of the optimized
negative-vacuum fits on $H_{\rm inf}$ requires separate calculations.

With the quantum cut removed, the independent hilltop solver
finds an accelerating $A_0$ solution
($w_{{\rm DE},0}\simeq-0.90$) at fixed $f/\mpl\simeq0.005$,
with $\chi^2\simeq1660.51$ and $\delta_i\simeq8\times10^{-80}$
($r\simeq6\times10^{-76}$).
Equation~\eqref{eq:profile_cut_mapping} shows that this point
would satisfy the cut only for
$H_{\rm inf}\lesssim6\times10^{-63}\,{\rm GeV}$,
placing it outside the domain of our AIC comparison.
For comparison, this initial angular displacement is about
twenty orders of magnitude smaller than
$H_0/(2\pi f)\simeq2\times10^{-59}$ evaluated at the same solution.

By Eq.~\eqref{eq:profile_q0}, present-day cosmic acceleration
($q_0<0$) requires
$3w_{{\rm DE},0}\Omega_{{\rm DE},0}<-(1+\Omega_{r0})$.
At the matter densities along the locally optimized $A_0$ branch,
the acceleration boundary $q_0=0$ corresponds to
$w_{{\rm DE},0}\simeq-0.49$.
Along this branch, $q_0$ changes sign between $r=10^3$ and
$r=3\times10^3$, corresponding to quantum-cut boundary scales
of order $10^{16}\,{\rm GeV}$
[Eq.~\eqref{eq:profile_cut_mapping}].
Both the refined local best-fit $A_0$ point and the locally
optimized point at $r\simeq4.7$ describe a decelerating universe today.
The $A_0$ posterior, however, predominantly favors accelerated
expansion: the median is $q_0=-0.394$, the central 68\% interval
is $[-0.458,-0.306]$, and only $0.16\%$ of the posterior weight
lies at $q_0>0$.

For reference, the smallest saved-chain $\chi^2$ values give
the diagnostic scores
$\chi^2_{\rm chain}+2N_{\rm par}-\widehat{\rm AIC}_L
\simeq-9.75$, $-8.16$, and $-8.41$ for $A_0$, $A_-$, and
$A_-^{\rm sub}$, respectively.
These comparisons use the local $L$ reference and are based
on saved samples, so they are not fully optimized AIC differences.
The local profile analysis described above applies only to $A_0$.

The two-unit offset between the AIC upper bounds for $A_-$
and $A_-^{\rm sub}$ and the $A_0$ estimate reflects the penalty
for the additional fitted parameter $\OmL$ at the shared
zero-vacuum point.
In terms of the global minima, the comparison is
\begin{equation}
 \Delta{\rm AIC}_{X/A_0}
 =\chi^2_{X,\min}-\chi^2_{A_0,\min}+2,
 \label{eq:aic_zero_vacuum_comparison}
\end{equation}
where $X\in\{A_-,A_-^{\rm sub}\}$.
Equal minima therefore favor $A_0$ by two AIC units;
either negative-vacuum model must lower $\chi^2_{\min}$ by
more than two relative to $A_0$ to be preferred.
The available bounds do not establish this ranking.
Since $A_-$ and $A_-^{\rm sub}$ have the same parameter count
and nested domains, their fully optimized AICs obey
${\rm AIC}_{A_-}\leq{\rm AIC}_{A_-^{\rm sub}}$,
although the difference remains undetermined.
Further minimization with $\OmL$ free may tighten the bounds.
The broad-prior $A_-/A_0$ comparison holds the $f$ range fixed,
whereas the $A_-^{\rm sub}/A_0$ comparison also restricts this range.

The quantum cut can affect AIC by restricting the allowed
parameter domain, but prior-normalization factors do not enter
$\chi^2$ or AIC. Also, $\Delta{\rm AIC}$ is not a Gaussian significance.
An improved fit relative to flat $\Lambda$CDM does not by itself
establish a preference for a negative cosmological constant,
since axion dynamics can also improve the fit.

\subsection{Bayesian model comparison}
\label{subsec:modelcomparison}

We now compute Bayes factors $B_{X/Y}\equiv Z_X/Z_Y$ to compare
the axion models with flat $\Lambda$CDM and with one another.
Evidence averages the likelihood over the normalized prior,
whereas AIC uses its maximum and a parameter penalty.
The two comparisons are distinct, and neither is determined
by a posterior-density maximum.

Table~\ref{tab:evidence} collects the evidences under the common
likelihood. The $L$ reference is independent of axion-prior
choices, while $A_0$ retains the broad quantum-conditioned
priors for the six shared parameters. The restricted
$A_-^{\rm sub}$ case is discussed in Sec.~\ref{subsec:subplanck}.

\begin{table*}[t]
\caption{Evidences for the completed axion runs and the unchanged
flat-$\Lambda$CDM reference. The axion priors are specified in
Tables~\ref{tab:axion_models} and \ref{tab:axion_priors}.
Convention I uses each model's normalized quantum-conditioned
prior; Convention II additionally conditions on numerical
acceptance. Convention-I errors combine the PolyChord and
initialization-counting uncertainties as in
Sec.~\ref{subsec:polychord}.}
\label{tab:evidence}
\begin{ruledtabular}
\begin{tabular}{lcccc}
Model & $D$ & $f_{{\rm valid},X}$ & $\ln Z^{\rm II}$ & $\ln Z^{\rm I}$ \\
\hline
$L$: flat $\Lambda$CDM & 4 & $0.126$ & $-849.608\pm0.111$ & $-851.678\pm0.115$ \\
$A_-$: broad quantum prior & 7 & $0.00292756$ & $-846.203\pm0.118$ & $-852.037\pm0.120$ \\
$A_0$: broad quantum prior, $\OmL=0$ & 6 & $0.00778531$ & $-846.704\pm0.121$ & $-851.560\pm0.123$ \\
$A_-^{\rm sub}$: quantum prior, $f\leq\mpl$ & 7 & $0.000908639$ & $-846.130\pm0.122$ & $-853.134\pm0.124$ \\
\end{tabular}
\end{ruledtabular}
\end{table*}

The $A_-$ initialization acceptance is
\begin{align}
 f_{{\rm valid},A_-}&=\frac{2000}{683163}=0.002927559,
 \nonumber\\
 \ln f_{{\rm valid},A_-}&=-5.833586.
 \label{eq:qprior_fvalid_result}
\end{align}
Using the full-precision output in Eqs.~\eqref{eq:evidenceI}
and \eqref{eq:evidenceII}, we obtain
\begin{align}
 \ln Z_{A_-}^{\rm II}&=-846.203378\pm0.118007,
 \nonumber\\
 \ln Z_{A_-}^{\rm I}&=-852.036964\pm0.120101.
 \label{eq:qprior_evidence_result}
\end{align}
The Convention-I uncertainty combines the PolyChord error
with the initialization-counting contribution
$\sigma(\ln f_{{\rm valid},A_-})\simeq0.02233$
in quadrature.

The accepted fraction is about $2.34\%$ of the calibration-only
value $f_{\rm cal}=0.125$ [Eq.~\eqref{eq:fcal}].
The reduction reflects background-branch restrictions,
bounds on the DE fraction at recombination and baryon drag
[Eq.~\eqref{eq:earlydecut}], and the flatness-shooting and
numerical acceptance criteria.
It is measured within the quantum-conditioned prior and is
distinct from the geometric quantum fraction $q_{A_-}$ in
Eq.~\eqref{eq:quantum_volume}.

For orientation, we adopt the conventional, heuristic scale
$|\ln B|<1$ (inconclusive), $1$--$2.5$ (weak), $2.5$--$5$
(moderate), and $>5$ (strong) \cite{Trotta:2008pf}.
The sign identifies the model with the larger evidence;
the magnitude determines the strength assigned on this scale.
Comparison with the unchanged $L$ reference gives
\begin{align}
 \ln B_{A_-/L}^{\rm I}&=-0.359\pm0.166,\\
 \ln B_{A_-/L}^{\rm II}&=3.405\pm0.162,
 \label{eq:qprior_bayes_result}
\end{align}
corresponding to $B_{A_-/L}^{\rm I}=0.70$ and
$B_{A_-/L}^{\rm II}=30.1$. The quoted errors combine the
independent log-evidence uncertainties of the two runs in
quadrature. Convention I is inconclusive, whereas
Convention II gives moderate support for $A_-$ over $L$ on
the stated scale.

Using $Z_X^{\rm I}=p_X Z_X^{\rm II}$, we obtain
\begin{equation}
 B_{A_-/L}^{\rm I}
 =\frac{p_{A_-}}{p_L}B_{A_-/L}^{\rm II},
 \label{eq:qprior_relative_acceptance}
\end{equation}
where the ratio of acceptance probabilities is estimated as
$p_{A_-}/p_L\simeq
f_{{\rm valid},A_-}/f_{{\rm valid},L}\simeq0.0232$.
The common calibration factor $f_{\rm cal}=0.125$ is included
in both acceptance probabilities and cancels in their ratio.
The larger Bayes factor in Convention II therefore results
from normalizing each prior over its accepted region.
Both conventions use the same observational data and likelihood.

Comparisons with $L$ change both the dynamical DE sector and
the vacuum term. To isolate the role of a negative cosmological
constant, we compare $A_-$ and $A_0$ using the same likelihood
and broad quantum-conditioned prior on their six shared parameters,
with $\OmL$ free in $A_-$ and fixed to zero in $A_0$.
With the settings of Eq.~\eqref{eq:polychordsettings}, the completed
six-parameter $A_0$ run has 17\,124 dead points and
$2.650488\times10^6$ likelihood evaluations, yielding
\begin{align}
 \ln Z_{A_0}^{\rm II}&=-846.704180\pm0.121402,
 \label{eq:A0_evidence_II}\\
 \ln Z_{A_0}^{\rm I}&=-851.559696\pm0.123428.
 \label{eq:A0_evidence}
\end{align}
Initialization accepts 2000 points from 256\,894 draws
from the normalized quantum-conditioned prior, giving
\begin{align}
 f_{{\rm valid},A_0}&=\frac{2000}{256894}
 \simeq0.00778531,
 \label{eq:A0_acceptance_fraction}\\
 \ln f_{{\rm valid},A_0}&\simeq-4.855516.
 \label{eq:A0_acceptance}
\end{align}
The Convention-I log evidence includes the correction
$\ln f_{{\rm valid},A_0}$.
Its quoted uncertainty combines the counting contribution
$0.022273$ with the PolyChord log-evidence uncertainty
$0.121402$ in quadrature, as described in
Sec.~\ref{subsec:polychord}.

Relative to the same $L$ reference, we obtain
\begin{align}
 \ln B_{A_0/L}^{\rm I}&=0.118\pm0.169,
 \label{eq:A0_vs_L_I}\\
 \ln B_{A_0/L}^{\rm II}&=2.904\pm0.164,
 \label{eq:A0_vs_L}
\end{align}
corresponding to $B_{A_0/L}^{\rm I}=1.13$ and
$B_{A_0/L}^{\rm II}=18.2$.
Convention I is inconclusive, whereas Convention II gives
moderate support for $A_0$ over $L$ on the adopted scale.
The direct comparison between the negative- and zero-vacuum
axion models gives
\begin{align}
 \ln B_{A_-/A_0}^{\rm I}&=-0.477\pm0.172,
 \label{eq:Aminus_vs_A0_I}\\
 \ln B_{A_-/A_0}^{\rm II}&=0.501\pm0.169,
 \label{eq:Aminus_vs_A0}
\end{align}
or $B_{A_-/A_0}^{\rm I}=0.620$ and
$B_{A_-/A_0}^{\rm II}=1.65$. The quoted uncertainties combine
the corresponding independent log-evidence errors in quadrature.
Both comparisons satisfy $|\ln B|<1$ and are therefore
inconclusive on the adopted scale; neither establishes a
preference between $A_-$ and $A_0$. The relation
\begin{equation}
 B_{A_-/A_0}^{\rm I}
 =\frac{p_{A_-}}{p_{A_0}}B_{A_-/A_0}^{\rm II},
 \label{eq:Aminus_A0_acceptance}
\end{equation}
shows that the change in the sign of $\ln B$ reflects the
different acceptance probabilities and the resulting prior
normalization.
Neither comparison provides compelling evidence for a negative
cosmological constant.

Reference~\cite{Luu:2025ade} compared peak posterior densities
at fixed $f=\mpl$. Our comparison uses the Bayesian evidence,
which integrates the likelihood over the normalized joint prior,
allowing both $f$ and $\theta_i$ to vary.
As Fig.~\ref{fig:fig4_compare} shows, solutions with the same
present-day DE density and pressure can have different earlier
expansion histories. Their contributions to the evidence depend
on both their fit to the data and the prior probability assigned
to the corresponding parameter regions.

\subsection{Sub-Planckian decay-constant benchmark}
\label{subsec:subplanck}

The $A_-^{\rm sub}$ model differs from $A_-$ only in the upper
prior bound on $f$, which is reduced to $\mpl$.
We renormalize the quantum-conditioned prior over this restricted
domain, keeping the likelihood and acceptance criteria unchanged.
This restriction is a theory-motivated benchmark
\cite{Shiu:2026bae}, rather than a universal ultraviolet bound.
Using the settings in Eq.~\eqref{eq:polychordsettings}, the run
terminates with 17\,222 dead points, 17\,169 posterior points,
and $2.141290\times10^6$ likelihood evaluations.
The weighted chain export contains 14\,645 rows and has an
effective sample size of $5407$ computed from the weights alone,
with the interpretation discussed in Sec.~\ref{subsec:posterior}.

The evidences are
\begin{align}
 \ln Z_{A_-^{\rm sub}}^{\rm II}
 &=-846.130162\pm0.122272,
 \label{eq:subplanck_evidence_II}\\
 \ln Z_{A_-^{\rm sub}}^{\rm I}
 &=-853.133724\pm0.124298.
 \label{eq:subplanck_evidence}
\end{align}
The initialization acceptance is
\begin{align}
 \ln f_{{\rm valid},A_-^{\rm sub}}&=-7.003562,
 \label{eq:subplanck_log_acceptance}\\
 f_{{\rm valid},A_-^{\rm sub}}
 &=\frac{2000}{2201093}=9.08639\times10^{-4}.
 \label{eq:subplanck_acceptance}
\end{align}
Following Sec.~\ref{subsec:polychord}, the Convention-I error
includes $\sigma(\ln f_{{\rm valid},A_-^{\rm sub}})\simeq0.02235$
in quadrature. The geometric quantum fraction
$q_{A_-^{\rm sub}}=0.999992185766$ is distinct from this acceptance
fraction and is already included in the prior transform.

Using the same $L$ reference, we obtain
\begin{align}
 \ln B_{A_-^{\rm sub}/L}^{\rm I}&=-1.456\pm0.169,
 \label{eq:subplanck_bayes_I}\\
 \ln B_{A_-^{\rm sub}/L}^{\rm II}&=3.478\pm0.165.
 \label{eq:subplanck_bayes}
\end{align}
On the adopted scale, Convention I gives weak support for $L$
over $A_-^{\rm sub}$ by a factor of approximately $4.3$,
whereas Convention II gives moderate support for $A_-^{\rm sub}$
over $L$ by a factor of $32.4$.
The difference reflects conditioning each prior on its accepted
parameter region in Convention II, as discussed in
Sec.~\ref{subsec:modelcomparison}.
Table~\ref{tab:fprior_bayes} compares these Bayes factors
with the broad-prior results.

\begin{table}[tbp]
\caption{Bayes factors relative to the unchanged flat-$\Lambda$CDM
reference for the two completed negative-vacuum decay-constant benchmarks.
Here $X=A_-$ and $A_-^{\rm sub}$ for the first and second rows,
respectively.
The joint prior is uniform in $\log_{10}(f/\mpl)$ and
$\theta_i$ over the domain defined by the stated decay-constant
range and the quantum condition. It is normalized separately
for each model.}
\label{tab:fprior_bayes}
\begin{ruledtabular}
\begin{tabular}{lcc}
Axion prior & $B^{\rm I}_{X/L}$ & $B^{\rm II}_{X/L}$ \\
\hline
$0.005\leq f/\mpl\leq10$ & $0.70$ & $30.1$ \\
$0.005\leq f/\mpl\leq1$ & $0.233$ & $32.4$ \\
\end{tabular}
\end{ruledtabular}
\end{table}

Comparing $A_-^{\rm sub}$ with the broad-prior model $A_-$,
we obtain
\begin{align}
 \ln\frac{Z_{A_-^{\rm sub}}^{\rm I}}{Z_{A_-}^{\rm I}}
 &=-1.097\pm0.173,
 \label{eq:subplanck_vs_broad_I}\\
 \ln\frac{Z_{A_-^{\rm sub}}^{\rm II}}{Z_{A_-}^{\rm II}}
 &=0.073\pm0.170.
 \label{eq:subplanck_vs_broad}
\end{align}
The corresponding Bayes factors are $0.334$ and $1.076$.
In Convention I, the central value lies just within the
weak-support range for $A_-$, but the quoted numerical
uncertainty extends into the inconclusive range.
Convention II is inconclusive, with a Bayes factor consistent
with unity within the quoted uncertainty.
The relation
\begin{align}
 \ln\frac{Z_{A_-^{\rm sub}}^{\rm I}}{Z_{A_-}^{\rm I}}
 &=\ln\frac{Z_{A_-^{\rm sub}}^{\rm II}}{Z_{A_-}^{\rm II}}
 +\ln\frac{p_{A_-^{\rm sub}}}{p_{A_-}}\nonumber\\
 &\simeq0.0732-1.1700
 \label{eq:subplanck_decomposition}
\end{align}
shows that the lower Convention-I evidence for $A_-^{\rm sub}$
primarily reflects its smaller accepted prior fraction.
The conditional evidences in Convention II are nearly equal.

We also check the restricted-prior evidence using the broad run.
For the region $R=\{f\leq\mpl\}$, we denote its probability
under the broad model's posterior by $P(R|\boldsymbol d)$
and under its normalized quantum-conditioned prior by $P(R)$.
Since the likelihood and acceptance criteria are unchanged,
the evidences satisfy
$Z_{A_-^{\rm sub}}^{\rm I}
=Z_{A_-}^{\rm I}P(R|\boldsymbol d)/P(R)$.
Using $P(R|\boldsymbol d)=0.331693$ and $P(R)=0.697063$,
we obtain $\ln Z_{A_-^{\rm sub}}^{\rm I}\simeq-852.780$,
which is $0.354$ above the dedicated-run estimate $-853.134$.
The corresponding Bayes factor is
$B_{A_-^{\rm sub}/L}^{\rm I}\simeq0.33$, compared with
$0.233$ from the dedicated run.
The reweighted estimate also gives
$\ln B_{A_-^{\rm sub}/A_-}^{\rm I}\simeq-0.743$,
which is inconclusive on the adopted scale.
The two quoted log-evidence errors combine to $0.173$ in
quadrature, but this omits uncertainty in $P(R|\boldsymbol d)$
and its covariance with the broad-run log evidence.
These errors alone therefore cannot establish the significance
of the discrepancy or identify which run is responsible.
We retain the evidence estimates from the dedicated
$A_-^{\rm sub}$ run. This unresolved discrepancy warrants
caution when interpreting small evidence differences between
the broad and restricted priors \cite{Higson:2019diagnostic}.

Comparing $A_-^{\rm sub}$ with the broad-prior zero-vacuum
model $A_0$, we obtain
\begin{align}
 \ln B_{A_-^{\rm sub}/A_0}^{\rm I}
 &=-1.574\pm0.175,\nonumber\\
 \ln B_{A_-^{\rm sub}/A_0}^{\rm II}
 &=0.574\pm0.172.
 \label{eq:subplanck_vs_A0}
\end{align}
The corresponding Bayes factors are $0.207$ and $1.78$.
On the adopted scale, Convention I gives weak support for
$A_0$, whereas Convention II is inconclusive.
This comparison changes both the allowed range of $f$ and
whether a negative cosmological constant is included.
To isolate the effect of the latter, we use the matched
broad-prior $A_-/A_0$ comparison in
Eqs.~\eqref{eq:Aminus_vs_A0_I} and \eqref{eq:Aminus_vs_A0}.
For the separate comparison with $L$, Table~\ref{tab:aic}
gives ${\rm AIC}_{A_-^{\rm sub}}\leq1674.741841593$,
below the local $L$ AIC estimate, whereas the Convention-I
evidence gives weak support for $L$.

For an initially frozen field,
$\rho_{{\rm DE},i}=U(\phi_i)-\Lambda^{(-)}$.
On the expanding branch,
$\dot\rho_{\rm DE}=-3H\dot\phi^2\leq0$,
so $\rho_{{\rm DE},0}\leq\rho_{{\rm DE},i}$.
Together with $U(\phi_i)\leq2m_\phi^2f^2$, this gives
\begin{align}
 2m_\phi^2f^2
 &\geq\rho_{{\rm DE},0}+\Lambda^{(-)}\nonumber\\
 &=3\mpl^2H_0^2
   (\Omega_{{\rm DE},0}+\OmL).
 \label{eq:subplanck_mass_scale}
\end{align}
At fixed $\Omega_{{\rm DE},0}$ and $\OmL$, a smaller $f$
therefore raises the lower bound on $m_\phi/H_0$.
A larger $m_\phi$ increases the magnitude of the hilltop
curvature, $|V_{,\phi\phi}(\pi f)|=m_\phi^2$,
so delaying the roll typically requires a smaller initial
displacement. The quantum cut imposes the competing condition
$\delta_i\geq\delta_{\min}$, with $\delta_{\min}\propto1/f$.
These requirements restrict the allowed hilltop histories,
and satisfying the energy bound alone does not guarantee
acceptance. Restricting the prior to sub-Planckian $f$
increases the relative prior weight of this constrained region
and can therefore reduce the accepted fraction.
This fraction also reflects calibration, branch, and numerical
exclusions, so it is not a measure of hilltop tuning alone.

\begin{figure*}[t]
\centering
\includegraphics[width=0.98\textwidth]{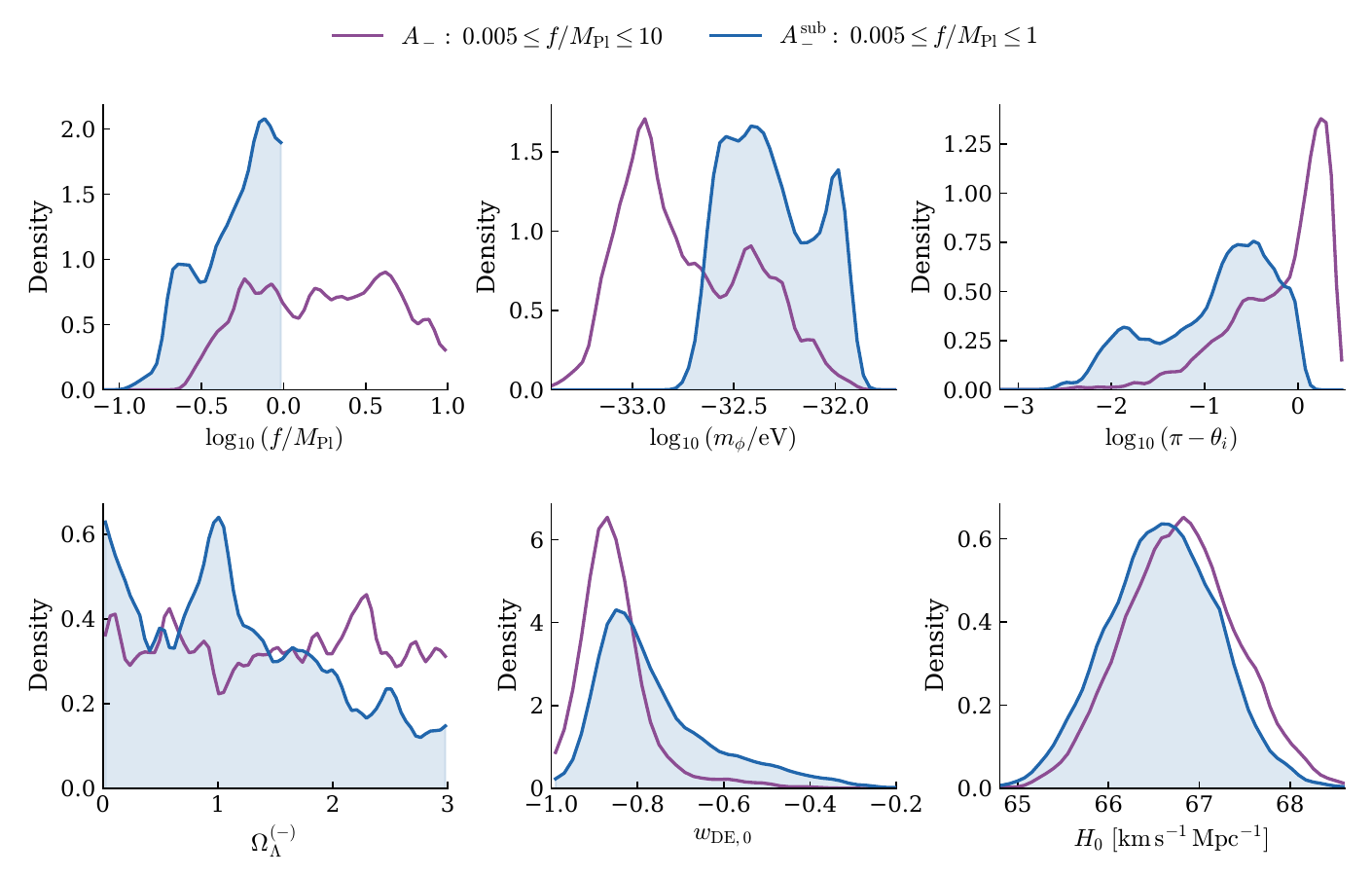}
\caption{Weighted one-dimensional marginal posterior densities for
the broad quantum-prior $A_{-}$ run (purple) and the dedicated
$A_{-}^{\rm sub}$ run (blue). Each curve is normalized using
the global weighted samples of the corresponding run.
The decay-constant upper bounds differ between the two runs,
as specified in the text; the marginals illustrate the resulting
prior dependence. The curves are weighted histograms smoothed
with a Gaussian kernel of one-bin width, whereas the tabulated
intervals are obtained from unsmoothed weighted quantiles.}
\label{fig:submarginals}
\end{figure*}

Figure~\ref{fig:submarginals} compares the one-dimensional
marginal posteriors of $A_-^{\rm sub}$ and $A_-$, while
Table~\ref{tab:subposterior} summarizes the restricted-prior
constraints. The central 68\% marginal credible intervals are
\begin{align}
 f/\mpl&\in[0.262,0.822],
 \label{eq:subposterior_f}\\
 m_\phi&\in[2.81,9.28]\times10^{-33}\,{\rm eV},\nonumber\\
 m_\phi/H_0&\in[1.97,6.56].
 \label{eq:subposterior_scales}
\end{align}
The equal-tail 95\% credible interval for $f/\mpl$ is
$[0.189,0.969]$. The one-sided 95\% posterior limits are
$f/\mpl>0.204$ and $m_\phi<1.10\times10^{-32}\,{\rm eV}$.

In Fig.~\ref{fig:submarginals}, the restricted-prior $f$
distribution has substantial density near its imposed upper
bound, while the mass distributions of both runs exhibit
multiple peaks. These features in the smoothed, finite-sample
distributions are not fully captured by means and central
credible intervals. In particular, Eq.~\eqref{eq:subposterior_f}
is conditional on the prior restriction $f\leq\mpl$ and does
not by itself establish an observational preference for a
sub-Planckian decay constant. Marginal credible intervals also
need not contain the parameter values at a likelihood maximum.

\begin{table}[tbp]
\caption{Posterior means and central 68\% credible intervals
for the dedicated quantum-prior $A_-^{\rm sub}$ run.
The interval convention and Hubble-constant units are the same
as in Table~\ref{tab:negativeLambdaPosterior}.}
\label{tab:subposterior}
\begin{ruledtabular}
\begin{tabular}{lcc}
Parameter & Mean & 68\% interval \\
\hline
$\log_{10}(m_\phi/{\rm eV})$ & $-32.313$ & $[-32.551,-32.032]$ \\
$\log_{10}(f/\mpl)$ & $-0.309$ & $[-0.581,-0.085]$ \\
$\log_{10}(\pi-\theta_i)$ & $-0.885$ & $[-1.667,-0.269]$ \\
$\OmL$ & $1.191$ & $[0.310,2.072]$ \\
$\theta_i$ & $2.874$ & $[2.603,3.120]$ \\
$H_0$ & $66.60$ & $[65.97,67.21]$ \\
$\Omega_{m0}$ & $0.3159$ & $[0.3098,0.3221]$ \\
$\sigma_8$ & $0.751$ & $[0.725,0.780]$ \\
$S_8$ & $0.770$ & $[0.745,0.799]$ \\
$w_{{\rm DE},0}$ & $-0.758$ & $[-0.877,-0.621]$ \\
\end{tabular}
\end{ruledtabular}
\end{table}

Restricting the prior shifts the posterior toward smaller $f$,
larger $m_\phi$, and smaller initial displacements from the hilltop.
The posterior standard deviations of $\log_{10}(m_\phi/{\rm eV})$
and $\log_{10}(f/\mpl)$ are $0.216$ and $0.217$, respectively,
whereas that of their sum is only $0.084$.
Regressing logarithmic mass on logarithmic decay constant gives
a slope of $-0.92$, close to the constant-product value $-1$
suggested by Eq.~\eqref{eq:mf_posterior_identity}.
Variations in the field angle and background quantities account
for departures from this relation.

The median initial angle is $\theta_i=2.959$, and the central
68\% credible interval for the hilltop displacement is
$\delta_i\in[0.0215,0.538]$.
All exported samples satisfy the quantum cut, with a minimum
ratio $\delta_i/\delta_{\min}(f)=22.7$.
This finite-sample gap does not establish how the likelihood
or posterior would change if the cut were relaxed.
The evidence remains conditional on $H_{\rm inf}$, which
determines both the allowed axion domain and its prior
normalization.

The present-day DE equation of state has posterior median
$w_{{\rm DE},0}=-0.798$ and mean $-0.758$, with central
68\% and 95\% credible intervals $[-0.877,-0.621]$ and
$[-0.934,-0.394]$, respectively.
The tail toward less negative values is more pronounced
than in the broad-prior run.
The posterior means are
$H_0=66.60\pm0.61\,{\rm km\,s^{-1}\,Mpc^{-1}}$ and
$S_8=0.770\pm0.029$, where the uncertainties are posterior
standard deviations; both remain close to the broad-prior
results.
The $\OmL$ posterior remains broad
(Table~\ref{tab:subposterior}).
As in the broad run, its positive equal-tail lower endpoint
does not by itself establish a preference over the
zero-vacuum model.

\begin{figure}[tbp]
\centering
\includegraphics[width=0.95\linewidth]{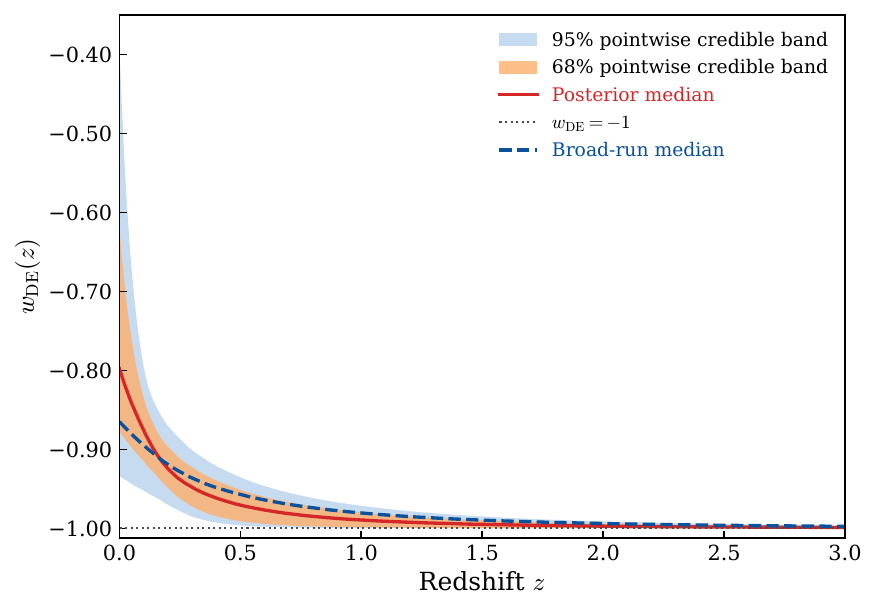}
\caption{Total-DE equation of state reconstructed from the
dedicated $A_-^{\rm sub}$ posterior. Light-blue and orange shading
denote central pointwise 95\% and 68\% credible intervals,
respectively, using the same band colors as
Fig.~\ref{fig:wdeposterior}. The red solid curve is the
pointwise sub-run median. The horizontal dotted line marks
$w_{\rm DE}=-1$, and the blue dashed curve gives the broad-run
median for comparison. The reconstruction uses 11269 weighted
histories at 301 redshifts in $0\leq z\leq3$.}
\label{fig:subwde}
\end{figure}

Figure~\ref{fig:subwde} reconstructs $w_{\rm DE}(z)$ using
the procedure in Sec.~\ref{subsec:wde}.
The posterior medians at $z=0.5$, $1$, and $3$ are
$-0.971$, $-0.989$, and $-0.9988$, respectively.
The central 68\% credible intervals at the first two redshifts
are $[-0.991,-0.951]$ and $[-0.999,-0.980]$.
Compared with the broad-prior reconstruction, the median
is less negative today but closer to $-1$ at $z=0.5$ and $1$.
This behavior is consistent with closer hilltop alignment
delaying the roll and larger masses driving more rapid
subsequent thawing.

\section{Discussion and conclusions}
\label{sec:conclusion}

\begin{figure}[tbp]
\centering
\includegraphics[width=0.95\linewidth]{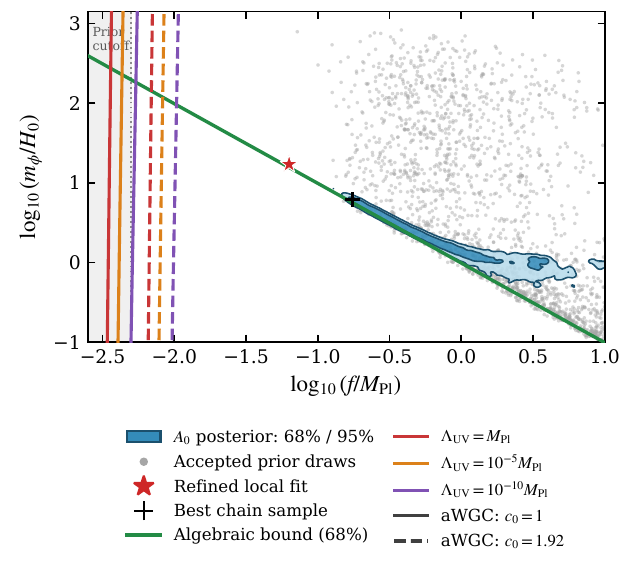}
\caption{$A_0$ posterior (dark/light blue: 68\%/95\%)
and 2000 accepted prior draws (gray), compared with the algebraic
bound (green, median and 68\% band) and aWGC curves
(solid: $c_0=1$; dashed: $1.92$). The star and cross mark the
refined local fit and best chain sample. Shading marks the
adopted prior exclusion $f/\mpl<0.005$. These necessary bounds do not determine
background viability or prior support; construction details
are given in the text.}
\label{fig:awgcA0}
\end{figure}

The DES Year 5 SN, DESI DR2, compressed-CMB, and corrected Gold-2017
growth data favor recently thawing histories within the negative-vacuum
axion model. With the broad prior, the median is
$w_{{\rm DE},0}=-0.865^{+0.068}_{-0.053}$. The central 68\%
intervals are $m_\phi\in[0.91,4.39]\times10^{-33}\,{\rm eV}$ and
$f/\mpl\in[0.589,4.899]$. The mass and decay constant are strongly
anticorrelated, while $\OmL$ remains weakly constrained. When the
upper prior bound on $f$ is lowered to $\mpl$, the central interval
becomes $f/\mpl\in[0.262,0.822]$. The posterior shifts toward
larger masses and initial angles closer to the hilltop. The field
then typically stays frozen longer and thaws more rapidly once
it starts to roll.

For Bayesian model comparison, we report Bayes factors under
the two prior conventions of Sec.~\ref{subsec:polychord}.
In Convention~I we average the likelihood over the normalized
quantum-conditioned prior and assign zero likelihood to rejected
points. The Bayes factors relative to $L$ are $0.70$ for $A_-$,
$1.13$ for $A_0$, and $0.233$ for $A_-^{\rm sub}$. The broad-prior
comparisons are inconclusive, while the restricted-prior comparison
weakly favors $L$. Conditioning on numerical acceptance in Convention~II
gives $30.1$, $18.2$, and $32.4$, respectively, all indicating
moderate support for the corresponding axion model over $L$
on the adopted scale. The observational likelihood is the same
in both conventions. The change follows from
the much smaller accepted fractions in the axion priors.

The matched broad-prior $A_-/A_0$ comparison directly tests the
negative cosmological constant. Its Bayes factors are $0.620$ and $1.65$
in Conventions~I and II, respectively. Both comparisons satisfy
$|\ln B|<1$ and are inconclusive on the adopted scale.
Comparing $A_-^{\rm sub}$ with the broad-prior $A_0$ also changes
the decay-constant range and therefore tests a different hypothesis.

Local likelihood optimization gives an estimated
$\widehat{\Delta{\rm AIC}}_{A_0/L}\simeq-10.87$, favoring
the zero-vacuum axion model $A_0$ over flat $\Lambda$CDM under
the AIC criterion. The $\chi^2$ improvement comes predominantly
from DES Year 5 SN, with an additional contribution from BAO.
The refined local best-fit point lies near the quantum-cut
boundary and describes a decelerating universe today
($q_0\simeq+0.577$), whereas most of the $A_0$ posterior favors
accelerated expansion (median $q_0=-0.394$).
The shared admissible point also gives
${\rm AIC}_X-\widehat{\rm AIC}_L\lesssim-8.868$ for
$X\in\{A_-,A_-^{\rm sub}\}$, relative to the local $L$ reference
(Table~\ref{tab:aic}). Global minima are not certified, and these
upper bounds do not determine either negative-vacuum model's
AIC ranking relative to $A_0$.

Along the explored $A_0$ branch, lowering the minimum displacement
by four decades to $r=10^{-4}$ yields no further likelihood gain
within the numerical resolution.
Raising $H_{\rm inf}$ to $4.7\times10^{13}\,{\rm GeV}$, near the
BK18 upper limit under the stated inflationary assumptions,
places the cut at $r\simeq4.7$ and increases $\chi^2$ and the
estimated AIC by only about $0.01$.
The lower $\chi^2$ found at $r\simeq6\times10^{-76}$ lies outside
the adopted quantum-conditioned domain.
These profile results apply only to the explored $A_0$ branch;
the $H_{\rm inf}$ dependence of the evidences and of the optimized
negative-vacuum fits requires separate calculations.

These conclusions depend on the adopted priors, including
$H_{\rm inf}$. They also depend on the fitting and rescaling
of the sound horizon, the compressed CMB likelihood, the
smooth-DE approximation for growth, and the RSD geometry
correction. The quoted evidence errors account for
nested-sampling uncertainty and, in Convention~I, the counting
uncertainty in the accepted prior fraction. Other sources of uncertainty
include likelihood approximations and modes missed by a finite
run \cite{Higson:2019diagnostic}. Reweighting the broad run to
$f\leq\mpl$ gives a Convention-I log evidence $0.354$ above the
dedicated-run result. Assessing this difference requires the
uncertainty in the reweighted posterior fraction and its covariance
with the broad-run evidence (Sec.~\ref{subsec:subplanck}). We
report the results of the dedicated run.

The $A_0$ samples also allow a quantitative comparison with
the axionic Weak Gravity Conjecture (aWGC). We define
$F\equiv f/\mpl$ and $\mu\equiv m_\phi/H_0$.
For the zero-vacuum cosine potential, its present value
$V_0\equiv V(\phi_0)$ obeys
\begin{align}
 V_0&=\frac{1-w_{{\rm DE},0}}{2}\rho_{{\rm DE},0}
       \leq2m_\phi^2f^2,\nonumber\\
 \mu F&\geq A_0^{\rm alg},\qquad
 A_0^{\rm alg}\equiv
 \sqrt{\frac34(1-w_{{\rm DE},0})\Omega_{{\rm DE},0}}.
 \label{eq:awgc_algebraic_bound}
\end{align}
The coefficient $A_0^{\rm alg}$ has posterior median $0.981$
and central 68\% credible interval $[0.956,0.999]$.
This algebraic condition underlies the static part of the
dynamical bound derived neglecting radiation in
Ref.~\cite{Shiu:2026bae}, but holds for either sign of
$\dd\epsilon/\dd N$.
Here $\epsilon\equiv-\dot H/H^2$ describes the late-time
expansion, and $\epsilon_0\equiv\epsilon(N=0)$ denotes its
present value.
For $A_0$, Eq.~\eqref{eq:mf_posterior_identity} gives
$(\mu F)^2\sin^2[\phi_0/(2f)]=(A_0^{\rm alg})^2$. Equality in
Eq.~\eqref{eq:awgc_algebraic_bound} is attained at the hilltop,
where the cosine potential reaches its maximum. All stored $A_0$
samples satisfy the algebraic bound, with minimum
$(\mu F/A_0^{\rm alg})^2=1.0000$ to the precision shown. This
checks the consistency of the reconstructed quantities.

The posterior weight with increasing $\epsilon$ today is
$25.3\%$. Of 400 histories drawn with posterior weights,
101 admit nontrivial intervals ending today with
$\epsilon(N)\leq\epsilon_0$.
For these histories, the time-dependent correction in Eq.~(5.1)
of Ref.~\cite{Shiu:2026bae} has a median of $1.9\%$ of the
static term and a maximum of $47\%$.
The largest correction raises the lower bound on $m_\phi$
by a factor $\sqrt{1+0.47}\simeq1.21$, or about $21\%$,
relative to the static bound.
All 400 histories satisfy the applicable inequality, using
the static condition for the remaining histories, with a
minimum left- to right-hand-side ratio of $1.0009$.

For the aWGC comparison, we make the additional assumption
that the instanton generating the dominant cosine term satisfies
$f S_{\rm inst}\leq c_0\mpl$.
The aWGC itself requires an instanton satisfying the bound,
which need not dominate the potential.
We parameterize the cosine amplitude as
$m_\phi^2f^2=\Lambda_{\rm UV}^4e^{-S_{\rm inst}}$.
Here $\Lambda_{\rm UV}$ sets the potential prefactor, while
$-\Lambda^{(-)}$ is an independent additive vacuum energy.
Under the stated assumptions, the aWGC implies
\begin{equation}
 m_\phi^2\geq\frac{\Lambda_{\rm UV}^4}{f^2}
 \exp\!\left(-\frac{c_0\mpl}{f}\right),
 \label{eq:awgc_mass_bound}
\end{equation}
as in Ref.~\cite{Shiu:2026bae}. The coefficient $c_0$ has benchmark values
$1.22$--$1.92$ in the four-dimensional wormhole-fragmentation
construction \cite{Lin:2026ztf}, conditional on its assumptions.

Figure~\ref{fig:awgcA0} compares Eqs.~\eqref{eq:awgc_algebraic_bound}
and \eqref{eq:awgc_mass_bound} with the weighted $A_0$ posterior.
We smooth the histogram in the two displayed logarithmic
coordinates with a Gaussian kernel of width 1.2 bins. We use the
unsmoothed weighted samples for all tabulated quantiles. The aWGC curves use
median $H_0=66.834\,{\rm km\,s^{-1}\,Mpc^{-1}}$, while
posterior coordinates use each sample's $H_0$. The aWGC-allowed
side is above each curve at fixed $F$, equivalently to the left
of its nearly vertical segment at fixed $\mu$. For $c_0=1$--$2$
and $\Lambda_{\rm UV}/\mpl=1$, $10^{-5}$, and $10^{-10}$,
intersections with the median algebraic-bound line lie around
$F=0.0036$--$0.011$ and $\mu=90$--$280$,
far from the main posterior ridge.
For $\Lambda_{\rm UV}=\mpl$ and $c_0=1$, the intersection
at $F\simeq0.0036$ lies below the adopted prior cutoff $F=0.005$.

Background viability additionally requires the field to supply the
present DE density after evolving from the allowed initial conditions.
The 2000 accepted prior draws illustrate the sampled region.
Their smallest decay constant, $F\simeq0.072$, exceeds the
admissible refined value $F\simeq0.063$. A 448-point search (14 decay constants,
16 masses, and $\delta_i=\delta_{\min}$ or $10\delta_{\min}$)
finds no accepted flat background at the sampled points with
$F\leq0.04$ or in the aWGC wedge, with the remaining cosmological
parameters fixed to
the refined solution and $50\leq H_0/({\rm km\,s^{-1}\,Mpc^{-1}})\leq85$.

With the other parameters free along $\delta_i=\delta_{\min}$,
an independent $A_0$ search finds a flat background at $F=0.045$
($H_0\simeq50.6\,{\rm km\,s^{-1}\,Mpc^{-1}}$) with a poor fit
($\Delta\chi^2\simeq1270$), and finds none at $F=0.040$.
If the axion energy is negligible today, flatness in $A_0$
requires $\Omega_{m0}\simeq1$ when radiation is neglected.
Using $\omega_m=\Omega_{m0}h^2$, this gives
$H_0\simeq100\sqrt{\omega_m}\,{\rm km\,s^{-1}\,Mpc^{-1}}$.
For $\omega_m\simeq0.140$, the required value is about
$37\,{\rm km\,s^{-1}\,Mpc^{-1}}$, below the allowed shooting range.
The lower edge of support remains undetermined by these finite
searches, which also leave open the possibility of admissible
points elsewhere in the aWGC wedge. The adopted cutoff $F=0.005$
is a prior choice. Its proximity
to the tip of the allowed wedge depends on $c_0$ and $\Lambda_{\rm UV}$.

For each posterior sample, the minimum coefficient required
by the assumed instanton interpretation is
\begin{equation}
 c_{\min}(\Lambda_{\rm UV})
 =\frac{f}{\mpl}\ln\!\left(
 \frac{\Lambda_{\rm UV}^4}{m_\phi^2f^2}\right).
 \label{eq:awgc_cmin}
\end{equation}
With our potential prefactor the logarithm in Eq.~\eqref{eq:awgc_cmin}
is the instanton action inferred from the sampled $m_\phi$ and $f$.
It is positive for the scales considered. The conjecture thus
requires $c_0\geq c_{\min}$ for each sample. Table~\ref{tab:awgc_cmin}
gives one-sided 95\% lower limits $55.3$, $46.1$, and $36.9$
at $\Lambda_{\rm UV}/\mpl=1$, $10^{-5}$, and $10^{-10}$.
These are fifth percentiles of the derived $c_{\min}$ posterior,
not central-95\% lower endpoints; $c_0$ is not a fitted parameter.
Even the refined local fit requires $c_{\min}=11.7$--$17.5$.

The tension with order-one $c_0$ is conditional on the model,
likelihoods, quantum prior, prefactor, and same-instanton
identification, not an exclusion of the aWGC generally.
Both lower limits and upper tails remain prior dependent;
a likelihood profile does not remove that dependence.
A different instanton could satisfy the aWGC while contributing
negligibly to the DE potential. Changing the potential prefactor
or adopting a multi-field construction can also alter the inference. For an underlying axion period
$2\pi f_a$ and positive integer instanton charge $k_{\rm inst}$,
the effective decay constant is $f=f_a/k_{\rm inst}$.
Increasing the charge of the same instanton does not provide
an independent escape at fixed $f$, since
$S_{\rm inst}f_a/(k_{\rm inst}\mpl)\leq c_0$ is identical to
$S_{\rm inst}f/\mpl\leq c_0$.
For $A_-$ and $A_-^{\rm sub}$ the algebraic bound becomes

\begin{equation}
 (\mu F)^2\geq\frac34\left[
 (1-w_{{\rm DE},0})\Omega_{{\rm DE},0}+2\OmL\right],
 \label{eq:awgc_negative_algebraic_bound}
\end{equation}
because $w_{\rm DE}$ includes the negative vacuum term.
For the same UV scales, Eq.~\eqref{eq:awgc_cmin} gives one-sided
95\% lower limits of $(109.3,91.1,72.9)$ for $A_-$ and
$(56.3,46.9,37.5)$ for $A_-^{\rm sub}$ from their respective
weighted production chains, in the order of Table~\ref{tab:awgc_cmin}.
Both negative-vacuum posteriors thus retain the conditional
tension with order-one $c_0$.
Testing the time-dependent bound requires the corresponding
background histories.

\begin{table}[tbp]

\caption{Derived $A_0$ constraints on $c_{\min}$ from
Eq.~\eqref{eq:awgc_cmin}. The fifth percentile is the
one-sided 95\% lower limit; the 2.5th percentile is the lower
endpoint of a central 95\% interval. The last column gives
the refined local likelihood point, not a posterior quantile.}
\label{tab:awgc_cmin}
\begin{ruledtabular}
\begin{tabular}{lrrrr}
$\Lambda_{\rm UV}/\mpl$ & 2.5\% & 5\% & Median & Local fit \\
\hline
$1$ & $51.1$ & $55.3$ & $214.8$ & $17.5$ \\
$10^{-5}$ & $42.6$ & $46.1$ & $179.1$ & $14.6$ \\
$10^{-10}$ & $34.1$ & $36.9$ & $143.4$ & $11.7$ \\
\end{tabular}
\end{ruledtabular}
\end{table}

Future measurements of galaxy clustering and weak lensing by
Euclid \cite{Euclid:2024overview}, together with Rubin/LSST
supernovae \cite{LSSTDESC:2018srd}, will provide complementary
probes of expansion and growth.
Within the axion models considered here, these data can tighten
constraints on the timing and rate of thawing and reduce
degeneracies among the axion parameters and the vacuum contribution.
Analyses using consistently normalized priors can then sharpen
comparisons with flat $\Lambda$CDM and clarify whether a negative
cosmological constant is supported relative to the matched
zero-vacuum axion model.

\acknowledgments
M.~C.~P. acknowledges support from JSPS KAKENHI Grant No.~26KF0150.
G.~S. is supported in part by DOE (HEP) Awards
DE-SC0017647 and DE-SC0023719, and a WARF Named Professorship, provided by the University of Wisconsin--Madison Office of the Vice Chancellor
for Research with funding from the Wisconsin Alumni Research Foundation. 
S.~T. acknowledges support from JSPS KAKENHI Grant Nos.~26K07090 and 26H00847 and from the Waseda University Special Research Projects (No.~2026C-486).

Anthropic Claude Code (Fable 5.1) and ChatGPT (OpenAI; GPT-6)
assisted with code inspection, selected chain and figure checks,
comparisons with analytic bounds, and manuscript revision. The authors supplied the manuscript, code,
and verification tasks, checked their suggestions against the code,
the stored PolyChord outputs, the analytic equations, and the cited literature,
and take responsibility for the final text, calculations, and
scientific conclusions.

\section*{Data availability}
The observational data used in this work are publicly available
from the sources cited in Sec.~\ref{sec:method}.
The weighted chains and numerical analysis code supporting our
results are available from the authors upon reasonable request.

\appendix
\section{Local likelihood-search settings}
\label{app:local_searches}

This appendix documents the local likelihood searches
underlying the AIC estimates and bounds in
Sec.~\ref{subsec:aiccomparison}, together with the numerical
checks of the $A_0$ profile.

We use the same physical prefilter, background acceptance tests,
and likelihood as in the production runs for Table~\ref{tab:aic}.
The axion fits also obey Eq.~\eqref{eq:quantum_cut}. For the
separate fixed-$r$ profile we allow initial displacements below
this threshold when $r<1$. Coordinate tolerances below refer
to the scaled parameters used by the optimizers.

For $L$, two bounded Nelder--Mead searches converge with
tolerances of $10^{-7}$ in both coordinates and $\chi^2$.
For $A_-$, Powell and Nelder--Mead searches start from 16 points:
a verified solution and low-$\chi^2$ broad-chain and cluster
samples. A Nelder--Mead refinement with a smaller initial
simplex and a $12\,000$-evaluation budget converges at
tolerances of $10^{-5}$ in coordinates and $10^{-4}$ in $\chi^2$.

For $A_0$, four bounded Nelder--Mead searches from distinct
low-$\chi^2$ chain samples converge, as does a smaller-simplex
refinement with tolerances of $10^{-5}$ in coordinates and
$10^{-4}$ in $\chi^2$.

For $A_-^{\rm sub}$, three bounded Nelder--Mead searches start
from the refined $A_0$ boundary point, the verified $A_-$ point
with $f<\mpl$, and the restricted chain's best sample, each
with a budget of 2000 likelihood evaluations. We refine the
best solution with a smaller simplex and a budget of 1500
evaluations. The tolerances
are $10^{-5}$ in coordinates and $10^{-4}$ in $\chi^2$;
logarithmic displacement from the quantum threshold enforces
the cut. The first two searches and refinement converge;
the chain-initialized search reaches its evaluation limit.

Freshly initialized models reproduce the final local values;
caching is disabled for the $A_0$ and $A_-$ checks.
An independent hilltop solver gives $\chi^2_L=1675.6100$, agreeing
within $10^{-3}$ in each sector. At the refined $A_0$ point its
default settings lower $\chi^2$ by $0.0045$. Matching the distance
integration, interpolation of background and growth observables,
growth initial conditions, and stored history resolution reproduces
each sector to within $3\times10^{-8}$ at both the refined and
best saved $A_0$ points. Both implementations give
$\widehat{\Delta{\rm AIC}}_{A_0/L}\simeq-10.87$.

For the fixed-$r$ $A_0$ profile, physical coordinates
$\log_{10}(f/\mpl)$ and $\log_{10}\delta_i$ permit evaluation
below the quantum cut. Setting $\delta_i=r\bar H_{\rm inf}/(2\pi f)$
leaves five free parameters $(\omega_b,\omega_c,\log_{10}(m_\phi/{\rm eV}),
\log_{10}(f/\mpl),\sigma_8)$. The 19 two-stage Nelder--Mead
searches use approximately 500--1600 likelihood calls each.
Starts from neighboring profile points converge; for $r\geq100$
we also start from the best saved chain point. The resulting
$0.001$--$0.003$ resolution characterizes the attained local
profile within the production implementation. The independent
six-point comparison uses the second solver's default numerical
settings.

\end{document}